\documentclass[aps,twocolumn,prl,twoside,amssymb,amsmath, superscriptaddress]{revtex4-2}
\usepackage[utf8]{inputenc}
\usepackage{bm}
\usepackage{multirow,amssymb,amsbsy,amsmath}
\usepackage{graphicx}
\usepackage{mathrsfs}
\usepackage{verbatim}
\usepackage{amssymb}
\usepackage{mathtools}
\usepackage{amsfonts}
\usepackage{bbm}
\usepackage{amsmath}
\usepackage{lmodern}
\usepackage{appendix}
\usepackage{soul}
\usepackage{ulem}

\usepackage{xcolor}

\definecolor{dark-gray}{rgb}{.35,.55,.55}
\definecolor{dark-blue}{rgb}{.0,.0,.6}
\usepackage[colorlinks=true,linkcolor=dark-blue,citecolor=dark-blue,urlcolor=dark-blue]{hyperref}

\newcommand{\tr}{{\rm Tr}}

\makeatletter
\usepackage{pifont}
\makeatother

\usepackage{dcolumn}            
\usepackage{epstopdf}                   
\usepackage{subfigure}

\newcommand{\bra}[1]{{\langle #1|}}
\newcommand{\ket}[1]{{|#1\rangle}}

\newcommand{\ketbra}[1]{{| #1\rangle \langle #1|}}

\newcommand{\be}{\begin{equation}}
\newcommand{\ee}{\end{equation}}
\newcommand{\bea}{\begin{eqnarray}}
\newcommand{\eea}{\end{eqnarray}}

\newcommand{\mean}[1]{\ensuremath{\langle{#1}\rangle}}

\newcommand{\eins}{\mathbbm{1}}
\newcommand{\WW}{\ensuremath{\mathcal{W}}}
\newcommand{\XX}{\ensuremath{\mathcal{X}}}

\newcommand{\kommentar}[1]{}
\newcommand{\forget}[1]{}

\begin{document}

\title{
Analyzing quantum entanglement with the Schmidt decomposition in operator space
}

\author{Chengjie Zhang}
\thanks{These two authors contributed equally to this work}
\email{chengjie.zhang@gmail.com}
\affiliation{School of Physical Science and Technology, Ningbo University, Ningbo, 315211, 
China}

\author{Sophia Denker}
\thanks{These two authors contributed equally to this work}
\email{chengjie.zhang@gmail.com}
\affiliation{Naturwissenschaftlich-Technische Fakult\"at, Universit\"at Siegen, Walter-Flex-Stra{\ss}e 3, 57068 Siegen, Germany}

\author{Ali Asadian}
\affiliation{Department of Physics, Institute for Advanced Studies in Basic Sciences (IASBS), Gava Zang, Zanjan 45137-66731, Iran}

\author{Otfried G\"uhne}
\email{otfried.guehne@uni-siegen.de}
\affiliation{Naturwissenschaftlich-Technische Fakult\"at, Universit\"at Siegen, Walter-Flex-Stra{\ss}e 3, 57068 Siegen, Germany}

\begin{abstract}
Characterizing entanglement is central for quantum information 
science. Special observables which indicate entanglement, 
so-called entanglement witnesses, are a widely used tool for 
this task. The construction of these witnesses typically relies 
on the observation that quantum states with a high fidelity to 
some entangled target state are entangled, too. We introduce a 
general method to construct entanglement witnesses based on the 
Schmidt decomposition of observables. The method works for two- 
and 
multi-particle systems and is strictly stronger 
than fidelity-based 
constructions. The resulting witnesses can also be used to 
quantify entanglement 
and to characterize its 
dimensionality. 
Finally, we present experimentally relevant 
examples, where our approach improves entanglement detection
significantly.
\end{abstract}
\date{\today}

\maketitle

{\it Introduction.---} 
In recent years, several experimental breakthroughs on 
different quantum technologies have been achieved. Examples are the 
demonstration of quantum supremacy with superconducting qubits 
\cite{arute}, the implementation of quantum cryptography using 
a satellite \cite{Pan} or in a device-independent 
manner \cite{nadlinger22, zhang22}, and the study of quantum phases using digital quantum 
simulation \cite{Lukin}. In such experiments large data sets are collected 
and the problem arises how to analyze them and connect them with the underlying 
quantum phenomena. For instance, if one wishes to reconstruct the density matrix 
of the quantum state arising in an experiment, methods like compressed 
sensing \cite{gross10}, matrix-product-state tomography \cite{cramer10}, shadow tomography \cite{huang20, nguyen22}, and forms of overlapping tomography \cite{cotler20, yu23} have been designed. 

For analyzing quantum correlations in experiments 
one frequently considers specific inequalities signaling 
the presence of correlations. The paradigmatic examples are Bell inequalities, 
whose violation signals the presence of quantum nonlocality \cite{brunnerbell}. 
Bell inequalities do not rely on assumptions on the measurement devices, 
and if knowledge about at least some of the implemented measurements 
is given, steering inequalities \cite{uolasteering} or entanglement witnesses \cite{rev4, rev5, rev6} are more efficient. In short, an entanglement witness is an 
observable with a nonnegative expectation value on all separable states, 
hence a negative expectation value signals the presence of entanglement. 
Clearly, finding all entanglement witnesses is a hard task, as it is 
equivalent to characterizing all entangled states, which is known
to be an NP-hard problem \cite{gharibian10}. Still, many constructions 
exist, often based on the idea of measuring the fidelity of the 
experimental state with some target state. If this fidelity is high {enough}, entanglement must be present.

In this paper, we present a method to analyze quantum entanglement
based on the so-called Schmidt decomposition of operators. The 
Schmidt decomposition is a ubiquitous tool when analyzing pure 
two-particle quantum states, but it can also be applied to 
bipartite observables. Our method leads to novel entanglement 
witnesses, which outperform fidelity-based witnesses and tolerate 
significantly more noise when analyzing multiparticle entanglement. 
Our approach is computationally simple and can also be used to 
quantify entanglement or its dimensionality.

{\it Entanglement and witnesses.---}
To start, recall that a bipartite quantum state $\varrho_{AB}$ 
shared by two parties, traditionally named Alice and Bob 
\cite{rev4, qibook} is separable if it can be written as  
$\varrho_{AB}= \sum_k p_k \ketbra{a_k} \otimes \ketbra{b_k}$, 
where the $p_k$ form a probability distribution. If a state can 
not be written in this way, it is entangled, 
which is, for many quantum tasks, a necessary condition to outperform 
classical protocols \cite{curty04, pezze09}. 
Unless stated otherwise, we assume that the 
dimensions of Alice's and Bob's space are the same, $d_A=d_B=d.$

For characterizing quantum entanglement, in experiments as well as in 
theory, entanglement  witnesses have turned out to be useful \cite{rev6, bourennane, gezaspin}, 
since they do not require full knowledge of the quantum state.
As already mentioned, entanglement witnesses have a positive expectation 
value on separable states, so measuring a negative expectation value
proves entanglement. For the construction of witnesses, several methods 
exist \cite{terhal, optimization, gezaspin, lurccnr, pianistates, dariusreview, HW, SIC}, and one of the well-known key methods are 
witnesses based on the fidelity with a given 
{\it pure} target state. They are of the form $\WW = \alpha \openone - \ketbra{\psi},$
where $\ket{\psi}$ is some pure entangled target state. This witness expresses
the fact that states with a high fidelity with $\ket{\psi}$, namely the ones
with $F_\psi = \bra{\psi} \varrho \ket{\psi} > \alpha$ are entangled, too. Three
remarks are in order. First, the coefficient $\alpha$ can directly be computed.
If $\ket{\psi} = \sum_{i=1}^R s_i \ket{ii}$ is the Schmidt 
decomposition (with decreasingly ordered Schmidt coefficients $s_i>0$ and Schmidt 
rank $R$), it is given by the maximal squared Schmidt coefficient
$\alpha = s_1^2$ \cite{bourennane}. Second, while fidelity-based witnesses are easy to 
construct, they have the disadvantage that they are not able to detect all 
entangled states,  
such as states with a positive partial transpose \cite{weilenmann20, faithful}. 
Still, fidelity-based witnesses 
have the advantage that they can be extended easily to the multiparticle 
case by considering the Schmidt decompositions for the different 
bipartitions \cite{bourennane}, this makes them the standard tool for analyzing entanglement in current experiments \cite{Nph1,Nph2,Nph3}.

{\it The main idea.---}
To introduce our main idea, let us start by pointing out 
the well-known fact that the Schmidt decomposition
does not apply to pure states only, but also to observables. 
Indeed, one can decompose any operator $\XX$ acting on a 
bipartite space in the operator Schmidt decomposition (OSD) 
\cite{oS,oS2}
\begin{equation}
\XX = \sum_{i=1}^S \mu_i G_i^A \otimes G_i^B.
\label{eq:osd}
\end{equation}
Here, the $\mu_i \geq 0$ are the operator Schmidt coefficients (OSC), chosen to be decreasingly ordered, and if $\XX$ is a quantum state, then
the largest one $\mu_1$ encodes the maximal correlation between 
two appropriately normalized observables. Further, the $G_i^A$ 
(resp. $G_i^B$) form an orthonormal basis of Alice's (Bob's) operator space. This means 
that $\tr(G_i^A G_j^A) = \delta_{ij}$, examples of such bases are 
the appropriately normalized Pauli or Gell-Mann matrices.
In fact, several works used the OSD, for example to analyze 
entanglement of mixed states \cite{cariello, cariello2, block, innsbruck}
or dynamics \cite{rescentOSD}.
We now write down our first main result, where we apply the OSD 
to a general operator $\XX$:
%

{\bf Observation 1.} 
Let $\XX$ be an operator with its OSD as in Eq.~(\ref{eq:osd}) and $\mu_1$ its largest OSC. Then
\begin{equation}
\WW = \mu_1 \openone - \XX
\label{eq:osdwitness}
\end{equation}
is an entanglement witness for bipartite entanglement.

Note that the choice of the parameter $\mu_1$
guarantees the positivity of the witness on  separable states, 
but for general $\XX$ there may be proper witnesses 
$\WW = \alpha \openone - \XX$ with a smaller $\alpha < \mu_1$.
This is in contrast to witnesses based on pure state fidelities, 
where $\alpha = s_1^2$ is optimal. In order to prove the observation, 
it suffices to show that the expectation 
value $\bra{a,b} \WW  \ket{a,b}$ is non-negative for an arbitrary 
pure product state $\ket{a,b}$; this implies the statement  for 
general separable states. First, writing $\XX$ in its OSD according 
to Eq.~(\ref{eq:osd}) it is clear that 
$\bra{a,b} \XX  \ket{a,b} \leq \mu_1 \sum_i  |\bra{a}G_i^A\ket{a} \bra{b}G_i^B\ket{b}|.$ 
Then, for $x=a$ and $x=b$ and any orthonormal basis of 
the operator space one has $\sum_i \bra{x}G_i^X\ket{x}^2 = 1$, 
this follows from the fact that for $\varrho=\ketbra{x}$ the relation 
$\tr(\varrho^2)=1$ holds. So, by the Cauchy-Schwarz inequality 
we have $\bra{a,b} \XX  \ket{a,b}  \leq \mu_1$ and the Observation 1 
follows.

From this simple construction of witnesses, several questions arise: How 
shall one choose the operator $\XX$ to detect a given entangled state 
$\varrho$? Which states can be detected by this construction? What about 
the characterization of high-dimensional entanglement? Can this  
method be extended to the multiparticle scenario, in order to detect
genuine multiparticle entanglement?

In the following, we will answer all these questions. For the moment, 
we would like to stress that the construction in Eq.~(\ref{eq:osdwitness})
contains the pure state fidelity-based witness mentioned in the second paragraph as a special
case, but still it is a more general description, so the OSD witnesses are strictly stronger. Indeed, starting from a 
pure state $\ket{\psi} = \sum_{i=1}^R s_i \ket{ii}$ one can directly calculate
the OSD of $\XX=\ketbra{\psi}.$ One finds $R^2$ nonzero operator Schmidt coefficients of the type $\{\mu_i\} =\{s_\alpha s_\beta\}$, and the largest 
one is hence given by $\mu_1 = s_1^2.$ Thus, the pure state fidelity-based witness is indeed a special case of Eq.~(\ref{eq:osdwitness}).

{\it Schmidt number witnesses.---}
Let us now explain how the method of OSD witnesses can be used to 
characterize the dimensionality of entanglement, as characterized 
by the 
Schmidt number. Given the Schmidt decomposition of a pure state 
as above, the number $R$ of nonzero Schmidt coefficients is called Schmidt 
rank, and is known to be an entanglement monotone characterizing the 
dimensionality of entanglement \cite{Eisert}. It can be generalized to mixed states
as follows. If a mixed state cannot be written as a convex decomposition 
into pure states with Schmidt rank $k$, then the mixed state has 
Schmidt number (SN) $k+1$ \cite{SN}. Note that in this classification, 
mixed states with SN one are just the separable states, while entangled states
have at least an SN of two.

Similar to entanglement witnesses, one can define Schmidt number witnesses as 
observables whose expectation values are positive for all states with SN $k-1$ 
such that a negative result indicates at least SN $k$ \cite{schmidtwit}. In our 
scheme, these witnesses may be constructed analogously to Eq.~(\ref{eq:osdwitness}), 
where the prefactor $\mu_1$ is replaced by a different number $\lambda_k$, which
is a not necessarily optimal bound on the overlap of pure Schmidt-rank $k-1$ states with the operator
$\XX$. It turns out that these $\lambda_k$ are simply given by the solution
of a $(k-1)$th order polynomial equation in the OSC of $\XX$. For example, for 
SN $k = 3$ we find $\lambda_3 = [\mu_1+\mu_4 + \sqrt{(\mu_1-\mu_4)^2+(\mu_2+\mu_3)^2}]/2$.
Then the witness $\WW = \lambda_3 \openone - \XX$ detects only three-dimensional
entanglement. However, for SNs greater than three, the prefactor is not so compact anymore, 
details on the computation of $\lambda_3$ and the prefactors for higher SN are given in Appendix A.
These witnesses can be seen as a generalization of the CCNR criterion for detecting Schmidt number (see also Observation 2 below), similar to
the one in Ref.~\cite{CCNRschmidt}. But one can choose $\XX$ such that it certifies the SN of states, for which the CCNR extension \cite{CCNRschmidt} fails, see also Appendix A.

{\it Estimating bipartite entanglement monotones.---}
{In many cases}, one is not only interested in detecting quantum entanglement,
{but also} wishes to quantify it and its resource character. For this
quantification, many entanglement monotones have been proposed \cite{ppt,ppt1,negativity,negativity21,CREN,EOFde1,2qubit1,2qubit2,concurrence3,concurrence4,concurrence5,Gour,Fan,Uhlmann,GME1,GME2,GME21}. A 
frequently used monotone is the concurrence \cite{EOFde1,2qubit1,2qubit2,concurrence3,concurrence4,concurrence5}, defined for pure states as 
$C(\ket{\psi}) = \sqrt{{2}[1-\tr(\varrho^2_A)]}$ and for mixed states via the 
so-called convex roof construction (see Appendix B for details). This 
quantity is notoriously difficult to compute, but with the OSD witness, 
it can be directly estimated. Indeed, one can show that 
\begin{equation}\label{Concurrence}
C(\varrho) \geq \sqrt{\frac{2}{d(d-1)}} (S-1)
\end{equation}
where $S= \max\{\tr(\varrho \XX)/\mu_1, 1\}.$ Note that 
Eq.~(\ref{Concurrence}) has a similar form of the result in 
Ref.~\cite{chen}, but interestingly 
analogous bounds can be derived for other measures, such as 
the convex-roof extended negativity \cite{CREN}, the G-concurrence \cite{Gour,Fan,Uhlmann} and  the geometric measure of entanglement  \cite{GME1,GME2,GME21}, details are given in Appendix B.
Moreover, they can finally be extended to the multiparticle case.

{\it Optimization of OSD witnesses.---}
Having established basic properties of the OSD witness, we 
can now ask how to choose the observable $\XX$ in an optimal 
manner. Consider an entangled state $\varrho$ that is detected 
by a witness as in Eq.~(\ref{eq:osdwitness}) with the $\XX$ as 
in Eq.~(\ref{eq:osd}). 
Since we want to minimize the expectation value of the witness we
can, without loss of generality, consider a witness where the 
the expectation values $\mean{G_i^A \otimes G_i^B}$ are positive
for the given state and the $\mu_i$ are positive. In addition, 
the witness may be renormalized to achieve ${\mu_1}=1$.
But then it is clear that the optimal choice of the other 
$\mu_i$ is to take $\mu_i=1,$ too.

%
So, an entangled quantum state is detected by a witness from 
Eq.~(\ref{eq:osdwitness}) if and only if it can be detected by a witness 
of the form ${\WW_{\mathrm{CCNR}}} = \openone - \sum_{i=1}^S G_i^A \otimes G_i^B$ 
These witnesses, however, are characteristic for the computable cross norm or 
realignment (CCNR) criterion \cite{CCNR1,CCNR2, lurccnr} and we have:

{\bf Observation {2}.} A bipartite quantum state can be detected by an OSD
witness as in Eq.~(\ref{eq:osdwitness}) if and only if it can be detected
by the CCNR criterion.

The critical reader may ask at this point, why we have defined
the OSD witnesses in the general form of  Eq.~(\ref{eq:osdwitness})
although the simpler subclass of CCNR witnesses contains all the 
relevant cases already. There are two reasons for that: First, 
as stressed above,
the direct connection to the CCNR criterion does not hold for witnesses 
for a higher Schmidt number. 
Second, the form  of the witness in 
Eq.~(\ref{eq:osdwitness}) is the key for the generalization to 
multiparticle entanglement. There, we will  search for multiparticle 
witnesses, which have the form as in Eq.~(\ref{eq:osdwitness}) for any 
bipartition. Restricting then the  attention to specific optimal 
witnesses for each bipartition does not lead to strong witnesses 
for the entire system \cite{Lancien} and there is a trade-off between the 
optimality of the bipartite witness and the efficiency for multiparticle 
entanglement detection.

So, let us discuss how a given OSD witness can be gradually optimized, 
this will be central for the discussion of multiparticle entanglement 
later. We consider an entangled state $\varrho$ (e.g., some pure state) 
which is affected by some separable noise $\sigma$ (e.g., the maximally 
mixed state $\openone/d^2$). So, the total state is of the form 
$\eta(p) = p \varrho + (1-p) \sigma$ and one can ask for the minimum of
the required visibility $p_{\rm crit}$,  such that all states with $p>p_{\rm crit}$ 
are detected by the OSD witness. 

A given OSD witness can be optimized in two directions. First, one may alter
the coefficients $\mu_i$ in Eq.~(\ref{eq:osd}), second one may change
the operators $G_i^X$ in the Schmidt decomposition. Let us first discuss
the optimization of the OSC. For a given OSD witness, 
one can directly compute the $p_{\rm crit}$ and, leaving all other 
quantities fixed, this is a function of the parameters $\{\mu_i\}$.
Then, one can compute the gradient 
$\nabla p_{\rm crit}(\{\mu_i\})$ and 
minimize $p_{\rm crit}$ 
with some steepest descent algorithm (see Appendix C for details). 
We stress that after adjusting the $\{\mu_i\}$ in one iteration step 
one can calculate the updated ${\tilde \mu}_1$ in order to guarantee 
that the updated $\tilde \WW$ is indeed a proper witness, so no 
fake entanglement detection can arise from this procedure. 


Second, we explain the optimization of the Schmidt operators $G_i^A$ while
keeping the $\{\mu_i\}$ and $\{G_i^B\}$ fixed. Since the $G_i^A$ form an
orthonormal basis, one can consider an infinitesimal rotation 
$G_i^A \rightarrow \tilde G_i^A=\sum_k O_{ik} G_k$ with $O_{ik}$ 
being an infinitesimal rotation matrix of the form
\begin{equation}
O = \eins + \sum_l \epsilon^{(l)} g^{(l)},   
\end{equation}
with $g^{(l)}$ being the generator matrices of the $SO(N)$.
Finally, one can write $p_{\rm crit}$  as a function of the 
$\epsilon^{(l)}$ and optimize the $G_i^A$ via a gradient algorithm.

In practice, these two approaches work very well, even if the initial
OSD witness was not chosen properly. For instance, for a bound entangled
state (the so-called UPB state in $3\times3$-systems) the procedures
directly find a witness that detects it, even if the initial witness was
not capable of detecting it. Details on the optimization procedures and
on the examples are given in Appendix C.

{\it Multiparticle entanglement.---}
Now we are ready to present the extension of OSD witnesses to the 
multiparticle case. Let us first recall the notion of genuine multiparticle 
entanglement (GME) \cite{rev5}. For the case of three particles, a 
pure state can be fully separable (e.g., $\ket{\psi^{\rm fs}}= \ket{000}$) 
or biseparable for some bipartition (e.g., $\ket{\psi^{\rm bs}} = \ket{\phi}_A \otimes \ket{\psi_-}_{BC}$,
where $\ket{\psi_-}$ is a two-qubit singlet state). 
Finally, a pure state is genuine multiparticle entangled 
if it is not biseparable with respect to any 
bipartition. Well-known examples of genuine multiparticle
entangled states for three qubits are the 
Greenberger-Horne-Zeilinger (GHZ) state $\ket{GHZ_3}= (\ket{000}+\ket{111})/\sqrt{2}$ 
and the W state $\ket{W_3}=(\ket{001}+\ket{010}+\ket{100})/\sqrt{3}.$ Similarly, 
one can define biseparability and genuine multiparticle
entanglement of more than three particles.

The generalization to mixed states goes via convex combinations. 
A mixed state is fully separable, if it can be written as a convex 
combination of pure fully separable states, that is $\varrho= \sum_k p_k \ketbra{\psi^{\rm fs}_k}.$ 
A state is biseparable 
if it can be expressed as a  convex combination of pure biseparable 
states, these pure states may be biseparable with respect to different 
bipartitions. Finally, mixed states are genuine multiparticle entangled, 
if they are not biseparable. 

For the characterization of genuine multiparticle entanglement, entanglement witnesses can be directly
used again and a witness for GME is defined by the property that it is non-negative on all biseparable states. The method of OSD witnesses can directly be
used to write down GME witnesses: Consider a tripartite
operator $\XX_{ABC}$. We can compute the OSD for the three bipartitions $A|BC$, $B|AC$
and $C|AB$, resulting in three maximal Schmidt coefficients $\mu_1^{A|BC}$, $\mu_1^{B|AC}$,
and $\mu_1^{C|AB}.$ 
Note that these are asymmetric scenarios for the OSDs, where the dimensions of the 
two sides are not the same.
Then, taking $\mu$ as the maximum 
of these, the operator
\begin{equation}
\WW = \mu \openone - \XX_{ABC}
\label{eq:mp-witness}
\end{equation}
has a positive expectation value on all pure
biseparable states, hence it is a witness for genuine
multiparticle entanglement; the generalization for more particles
is described in Appendix E.

It is clear that the witness in Eq.~(\ref{eq:mp-witness})
is more general than fidelity-based witnesses for multiparticle entanglement. Such fidelity-based 
witnesses have been a standard tool to analyze 
GME in experiments in the last years, so we
will analyze in the following the advantage occurring
from the construction in Eq.~(\ref{eq:mp-witness}).

{\it Examples of multiparticle states.---}
Now we are ready to use our methods to derive stronger 
witnesses for genuine multiparticle entanglement. In the 
following, we explain our approach for the three-qubit W 
state $\ket{W_3}$, the approach for other states is similar.
A known witness for genuine multiparticle entanglement in 
the vicinity of the W state is \cite{acinprl2001} 
\begin{equation}
\WW = \frac{2}{3} \openone  - \ketbra{W_3}. 
\label{eq:standardwwitness}
\end{equation}
This can be viewed as an OSD witness from
Eq.~(\ref{eq:mp-witness}) with $\XX_{ABC}=\ketbra{W_3}.$ 
The recipe for its improvement is as follows. We consider 
$\varrho=\ketbra{W_3}$ as entangled target state and wish 
to maximize the robustness for the separable noise given by 
$\sigma = \openone/8$. For a given bipartition one can 
improve $\XX_{ABC}$ by adjusting the Schmidt coefficients 
or the Schmidt operators as outlined above. We then go 
through the bipartitions and for each bipartition we 
improve the witness by a combination of the two 
optimization methods. Numerical details of the  
procedure are given in Appendix C.

The starting witness in Eq.~(\ref{eq:standardwwitness}) 
requires a visibility of $p_{\rm fid} \geq 0.620$ in order 
to detect GME. Already after some iterations of the optimization 
procedure one arrives at a witness for which the required visibility 
is reduced to $p_{\rm OSD} \geq 0.556$, demonstrating 
the superiority of the OSD witness over the fidelity-based 
construction.

We have applied the same method to a variety of other
multi-qubit states. This includes the three-qubit uniform
hypergraph state $\ket{H_3}$ \cite{1404.6492} and 
the four-qubit W state $\ket{W_4}$, Dicke state $\ket{D_4}$ 
and singlet-state $\ket{\Psi_4}$ \cite{rev5}. For all these
states we found a significantly improved noise robustness,
see Table~\ref{tab:mp-results} for concrete values. Detailed 
forms of the states as well as the results for other states 
are given in Appendix C. Note that OSD witnesses do not 
improve the fidelity-based witness $\WW = {\openone}/{2} - \ketbra{GHZ_3}$
for the GHZ state, as this witness is known to be optimal for 
maximally mixed noise \cite{bounds15}.

\begin{table}[t]
    \centering
    \begin{tabular}{|c|c|c|}
    \hline
       state  & visibility $p_{\rm fid}$  &  visibility $p_{\rm OSD}$ \\
       \hline
        $\ket{W_3}$ & $13/21 \approx {{0.619}}$ & 0.556
         \\
         \hline
          $\ket{H_3}$ & $5/7 \approx 0.714$ & 0.545
         \\
         \hline
          $\ket{W_4}$ & $11/15 \approx 0.733$  & 0.714
         \\
         \hline
          $\ket{D_4}$ & $29/45 \approx {0.644}$ & 0.540
         \\
         \hline
          $\ket{\Psi_4}$ & $11/15 \approx 0.733$ & 0.572
         \\
         \hline
    \end{tabular}
    \caption{Improvement of the noise robustness of entanglement detection for various multi-qubit states. For five different states, the required visibility $p_{\rm fid}$ for the fidelity-based witness and $p_{\rm OSD}$ for the OSD witness are shown. Look at the text for further details.}
    \label{tab:mp-results}
\end{table}
{\it Analytical approaches.---}
Two analytical approaches are worth to be mentioned. First, it is 
also possible to construct the OSD witnesses analytically
by starting from a pure high-dimensional quantum state and interpreting
this as an operator on a lower-dimensional space. For instance, for the 
GHZ state $\ket{GHZ} = (1/2) \sum_{i=1}^4 \ket{iii}$ on three four-level
systems the vector Schmidt decomposition is directly given. Consequently, 
taking $\XX = \sum_{i=1}^4 \mathcal{G}_i^A\otimes \mathcal{G}_i^B \otimes \mathcal{G}_i^C$ for arbitrary orthonormal bases $\mathcal{G}_i^X$, $(X = A,B,C)$ on three qubits will always result 
in an entanglement witness $\mathcal{W}= \openone - \XX.$ This ansatz can 
be generalized using arbitrary highly entangled pure states. Most importantly, 
given such a witness with a fixed structure, one can optimize the operators 
$\mathcal{G}_i^X$ for given states by an 
iteration of purely analytical steps, 
which is indeed more general than a simple
optimization over local unitary transformations. 
Details and examples are given in Appendix~D.
Second, note that the operators $\mathcal{G}_i^X$ are actually local orthogonal 
observables. Thus we can use the results from Ref.~\cite{chengjieCV}
to define the OSD witnesses for continuous variable systems, too.



{\it Multipartite entanglement measures.---}
Again, the novel multiparticle witnesses can be made 
quantitative and be used to estimate monotones for 
genuine multiparticle entanglement. One possibility 
to build such monotones is to start with an entanglement
monotone $E$ for pure two-particle states. Then, one can define for a multiparticle state the global entanglement $E_{\rm GME}$ as the minimum of $E$ for all bipartitions. Finally, 
one extends this to mixed states via the convex roof construction. 

Such entanglement monotones can be directly estimated from
the expectation value of the witness in Eq.~(\ref{eq:mp-witness}). 
For instance, one may consider the multiparticle
version of the concurrence. Then, one finds
$
C_{\rm GME} \geq \sqrt{{2}/{[m(m-1)]}} (S-1)
$
where $S= \max\{\tr(\varrho \XX_{ABC})/\mu, 1\}$
and $m$ is, for the special case of tripartite systems the
maximum of dimensions of Alice, Bob, and Charlie. This approach 
can be generalized to other measures and more particles, 
details are given in Appendix E.

{\it Conclusion.---}
We have introduced a novel method to characterize entanglement for quantum systems of two 
or more particles. The resulting entanglement witnesses 
are strictly stronger than the widely used fidelity-based 
witness and can improve entanglement detection in realistic
scenarios significantly. On the technical level, the approach does not involve advanced numerical tools such as semidefinite programming. The method can be seen as an extension of the CCNR criterion of separability to the 
multiparticle case, in the same sense as 
Ref.~\cite{Jungnitsch} presented an extension of the PPT criterion to the multiparticle case.

Several new lines of research emerge from our findings. First, it would be highly 
desirable to further characterize the resulting witnesses analytically for interesting families of quantum states.  Second, entanglement witnesses can 
also be used to characterize other properties of quantum states, such as the teleportation fidelity \cite{teleportation}, the 
distillability \cite{distillability}, and the multipartite Schmidt vector \cite{Huber, SVec} so it is relevant to apply our methods to these cases.
Third, the statistical analysis of entanglement tests from finite data has 
become essential in the last years \cite{flammia}, so our approaches also need to be 
analyzed from this viewpoint. Fourth, thinking of experimental implementations, it is desirable to give an estimation of the errors occurring 
when assuming small deviations of the desired measurements. In fact, for a 
special case of two-particle OSD witnesses this was recently discussed \cite{morelli}, but it remains open to generalize this approach further 
and apply it to the multipartite case.
Finally, in general, as fidelity-based entanglement witnesses 
have been used for many experiments, the presented improvement may allow 
for novel and exciting experiments. 

We thank {Jonathan Steinberg} 
for discussions. This work is partially supported by  the Innovation Program for Quantum Science and Technology (Grant No. 2021ZD0301200), the National Natural Science Foundation of China (Grant No. 11734015), and K. C. Wong Magna Fund in Ningbo University. S.D. and O.G. are supported by the Deutsche Forschungsgemeinschaft (DFG, German Research Foundation, project numbers 447948357 and 440958198), the Sino-German Center for Research Promotion (Project M-0294), the ERC (Consolidator Grant 683107/TempoQ), and the German 
Ministry of Education and Research (Project QuKuK, BMBF Grant No. 
16KIS1618K). S.D. is further supported by the House of Young Talents of
the University of Siegen. A.A. acknowledges the support from DAAD fund for bilateral project.

\newpage








\onecolumngrid
\appendix
\newpage
\section{Appendix A: Schmidt number witnesses}
In this appendix, we discuss how OSD witnesses can be used to analyze the dimensionality 
of entanglement by characterizing the Schmidt number (SN) of quantum states. First, we explain 
in detail the derivation of witnesses for SN three, as mentioned in the main 
text. Then, we derive a construction for higher SNs. Finally, we present two
examples which demonstrate the power of our methods. First, we present a quantum state 
whose SN can not be certified with fidelity-based witnesses, but with
OSD witnesses. Second we present an example which shows that the OSD witnesses for SN are
stronger than simple extensions of the CCNR criterion.

\subsection{A1:  Witnesses for SN three}
We can use our scheme to construct SN witnesses according to 
$\WW_{k} = \lambda_k\openone-\XX$. In order to compute the 
coefficient $\lambda_k$ such that $\WW_{k}$ is indeed a SN 
witness one has to maximize the expression 
$\bra{\psi_{k-1}}\XX \ket{\psi_{k-1}} =  \tr(\XX \ketbra{\psi_{k-1}})$ 
with respect to all pure states $\ket{\psi_{k-1}}$ with Schmidt 
rank $R =k-1$. This can be simplified by considering an upper bound: $\tr(\XX \ketbra{\psi_{k-1}})\leq \sum_{ij} \mu_i \tilde{\mu}_{j}$, where the $\mu_i$ and $\tilde{\mu}_j$ are the OSCs of $\XX$ and $\ketbra{\psi_{k-1}}$, respectively. Then, for $k=3$ the maximum of this expression can be found analytically, which results in maximizing a second order polynomial as mentioned in the main text; now we explain
the details.

For $k = 3$ the decreasingly ordered OSCs $\tilde\mu_{j}$ of the projector 
$\ket{\psi_2}\bra{\psi_2}$ are given by the products of the vector 
Schmidt coefficients of the states $\ket{\psi_2}$, that is 
$\{s_1s_1,s_1s_2,s_2s_1,s_2s_2\}$. The maximization then 
simplifies to 
\begin{align}
\max_{\tilde{\mu}_j} \sum_{ij} \mu_i \tilde{\mu}_{j} = 
\max_{s_1,s_2}(\mu_1 s_1^2 + \mu_2 s_1 s_2 + \mu_3 s_2 s_1 + \mu_4 s_2^2) 
= \max_{s_1,s_2}f(s_1,s_2),
\end{align}
where the coefficients $s_1$ and $s_2$ obey the constraints $s_1^2+s_2^2=1$, $s_1,s_2>0$ 
and $s_1 \geq s_2$. Taking a closer look at the function $f(s_1,s_2)$, one finds that 
it can also be written as a matrix vector multiplication
\begin{align}
f(s_1,s_2) = \bra{s} M \ket{s}
\label{eq:eig}
\end{align}
with the normalized vector $\ket{s} = (s_1,s_2)^T$ and the matrix
\begin{align}
    M = \left(
    \begin{array}{rr}
        \mu_1 & \mu_2 \\
        \mu_3 & \mu_4\\
    \end{array}
    \right).
\end{align}
Since the matrix $M$ is not necessarily symmetric, it needs to be symmetrized, such 
that Eq.~(\ref{eq:eig}) describes an eigenvalue problem, which is then solved by the 
largest eigenvalue of the matrix $M^\mathrm{symm} = {(M + M^T)}/{2}$:
\begin{align}
\max_\ket{s} \bra{s}M\ket{s} = \max_\ket{s}\bra{s}M^\mathrm{symm}\ket{s} 
\label{eq:eigval}.
\end{align}
Then, the eigenvalues $\gamma$ can be computed, which for the $2\times 2$ 
matrix $M^\mathrm{symm}$ are given by:
\begin{align}
\gamma_{1,2}  = \frac{1}{2}\left(\mu_1+\mu_4\pm \sqrt{\mu_{14}^2+\mu_{23}^2}\right),
\end{align}
where we used the short-hand notations $\mu_{14} := \mu_1 -\mu_4$ and 
$\mu_{23} := \mu_2 + \mu_3$. Since all $\mu_i$ as well as $\mu_{23}$ 
and $\mu_{14}$ are positive, the maximum eigenvalue is
\begin{align}
\lambda_3 = 
\gamma_\mathrm{max} = 
\frac{1}{2}
\left(\mu_1+\mu_4+ \sqrt{\mu_{14}^2+\mu_{23}^2}\right).
\label{eq:k3}
\end{align}
 Finally, we note that this value can indeed be reached with the given constraints 
 on the $s_i$. The coefficients of the eigenvector $\ket{s}$ corresponding to the 
 maximum eigenvalue of $M^\mathrm{symm}$ are all positive, since having a negative 
 coefficient would make the expression $\bra{s}M^\mathrm{symm}\ket{s}$ smaller. 
 Moreover, the eigenvector is normalized and therefore $s_1^2+s_2^2=1$ is fulfilled. 
 The last condition the $s_i$ need to fulfill is that they should be ordered decreasingly (
 $s_1\geq s_2$). This, however is guaranteed by the way the matrix $M$ and therefore $M^\mathrm{symm}$ is constructed. Since the OSCs $\mu_i$ are also in decreasing order 
 and therefore $\mu_1 \geq \mu_4$, the first entry $s_1$  of the eigenvector maximizing 
 Eq.~(\ref{eq:eigval}) must be greater or equal to the second one, $s_1 \geq s_2$.

\subsection{A2: Witnesses for higher SN}

Now we consider witnesses certifiying SN $k = 4$. Therefore we want to find 
the coefficient $\lambda_4$ such that the Schmidt witness has a non-negative
expectation value on all states $\ket{\psi_3}$ with SN $k-1=3$. Proceeding 
analogously to the case $k=3$, the OSCs of the projector $\ket{\psi_3}\bra{\psi_3}$ 
are given by 
$\{\tilde{\mu}_j\} = \{s_1s_1, s_1s_2, s_2s_1, s_1s_3, s_3s_1, s_2s_2, s_2s_3, s_3s_2, s_3s_3\}$. 
To obtain the function $f(s_1,s_2,s_3) = \sum_{ij}\mu_i\tilde{\mu}_{j}$, the OSCs 
$\tilde{\mu}_j$ have to be sorted decreasingly. In the case $k=4$, however, there
is no unique order: Depending  on the actual values of the $s_i$, one has either 
$s_1s_3 > s_2s_2$ or $s_1s_3 < s_2s_2$. They also might be equal ($s_1s_3 = s_2s_2$), 
but then the order would not matter. This leads to two possible functions $f(s_1,s_2,s_3)$ 
and $g(s_1,s_2,s_3)$ that have to be considered:
\begin{align}
f(s_1,s_2,s_3) & =\mu_1s_1^2+\mu_2s_1s_2+\mu_3s_2s_1+\mu_4\underline{s_1s_3}
+\mu_5\underline{s_3s_1}+\mu_6\underline{s_2^2}+\mu_7s_2s_3+\mu_8s_3s_2+\mu_9s_3^2,
\nonumber
\\
g(s_1,s_2,s_3)& =\mu_1s_1^2+\mu_2s_1s_2+\mu_3s_2s_1+\mu_4\underline{s_2^2}
+\mu_5\underline{s_1s_3}+\mu_6\underline{s_3s_1}+\mu_7s_2s_3+\mu_8s_3s_2+\mu_9s_3^2.
\end{align}
The coefficient $\lambda_4$ is then given by the maximum of those two functions: 
\begin{align}
\max_{s_1,s_2,s_3} \{f(s_1,s_2,s_3), g(s_1,s_2,s_3)\}.
\end{align}
This can be found by interpreting $f(s_1,s_2,s_3)$ and $g(s_1,s_2,s_3)$
in terms of matrices, analogously to the case $k =3$. The corresponding matrices 
are given by
\begin{align}
    M_1 =
    \left(
    \begin{array}{ccc}
         \mu_1&\mu_2&\boxed{\mu_4}  \\
         \mu_3&\boxed{\mu_6}&\mu_7 \\
         \boxed{\mu_5}&\mu_8&\mu_9\\
    \end{array}
    \right),
    \hspace{1cm}
    M_2 =
    \left(
    \begin{array}{ccc}
        \mu_1 &\mu_2&\boxed{\mu_5}  \\
        \mu_3 &\boxed{\mu_4}&\mu_7 \\
        \boxed{\mu_6} &\mu_8&\mu_9
    \end{array}
    \right).
    \label{eq:m2}
\end{align}
Proceeding as before, the coefficient $\lambda_4$ is given by
\begin{align}
\lambda_4 = \max \{ \mathrm{maxEig}(M_1^\mathrm{symm}), \mathrm{maxEig}(M_2^\mathrm{symm})\},
\end{align}
where $\mathrm{maxEig}(M_n^\mathrm{symm})$ describes the maximum eigenvalue of $M_n^\mathrm{symm}$ 
and $M_1^\mathrm{symm}$ and $M_2^\mathrm{symm}$ are the symmetrized matrices $M_1$ and $M_2$ in 
Eq.~(\ref{eq:m2}). Since the $M_n$ are $3\times 3$ matrices, $\lambda_4$ is the solution of a 
third order polynomial in the $\mu_i \,\,( i =1,...,9)$, which are the first nine OSCs of the 
operator $\XX$. So, computing $\lambda_4$ is numerically straightforward.

For SN higher than four, however, the computation is still straightforward but requires more 
effort. In these cases there are even more OSCs $s_{\alpha} s_{\beta}$ to arrange in decreasing 
order and therefore more possibilities to define the matrices $M_n^\mathrm{symm}$ occur. One 
can roughly estimate the number of matrices one has to check for a given SN $k$ using the concept 
of Young tableaux. A detailed discussion of this is given in Ref.~\cite{ma}.

\subsection{A3: Two relevant examples}

For our first example, we consider the mixed two-ququad state
\begin{align}
\varrho_3 = \frac{1}{2}\ket{\phi^3_+}\bra{\phi^3_+}+\frac{1}{4}(\ket{23}+\ket{32})(\bra{23}+\bra{32}),
\end{align}
with
\begin{equation}
\ket{\phi^3_+} = \frac{1}{\sqrt{3}}(\ket{00}+\ket{11}+\ket{22}),
\end{equation}
which has SN three. Fidelity-based Schmidt witnesses here only detect Schmidt 
number two \cite{weilenmann20}, which is equivalent to certifying entanglement. 
Using our scheme, however, we can construct a witness detecting the SN three
of this state.

As already mentioned in the main text, for bipartite entanglement witnesses 
the best choice of the operator $\XX$ is taking the Schmidt operators from 
the target state $\varrho_3$ and setting all OSCs to one. Thus it is a natural 
approach to choose the operator $\XX$ the same way for Schmidt witnesses. 
This yields $\lambda_3 = 2$ and consequently, the Schmidt witness for 
SN three is given by
\begin{align}
\mathcal{W}_{3} =  2\openone-\sum_i \tilde{G}_i^A \otimes \tilde{G}_i^B,
\end{align}
where the $\tilde{G}_i^{A/B}$ are the Schmidt operators of $\varrho_3$. This 
witness certifies SN three of the state $\varrho_3$ with visibility $p = 0.830$.

We add that for the special choice of $\XX = \sum_i \tilde{G}_i^A\otimes \tilde{G}_i^B$ 
the coefficient $\lambda_k$ reduces to $\lambda_k = k-1$. This is due to the fact that 
in this case there is only one possibility to arrange the OSCs in a matrix $M$, namely 
a  $(k-1)\times(k-1)$ matrix with all entries equal to one. These matrices have 
only one nonzero eigenvalue which is given by $k-1$. Note that this form of the witness 
corresponds to an extension of the  CCNR criterion. Indeed, as shown in Ref.~\cite{CCNRschmidt} 
the CCNR criterion can also be applied to characterize the dimensionality of entanglement; 
one has $\sum_i\mu_i\leq k$ if the state has SN $k$.

Still, with our second example we show that the OSD witnesses for the SN also detect 
states where the extended CCNR criterion fails. To do so, consider the pure two-ququad 
state
\begin{equation}
\ket{\psi_3} = \sqrt{1-2\varepsilon^2}\ket{00}+\varepsilon(\ket{11}+\ket{22}),
\end{equation}
where $\varepsilon = 0.1$. This state has SN three and is detected by the fidelity-based 
witness $\WW = (1-\varepsilon^2)\openone-\ketbra{\psi_3}$ \cite{faithful}, which is a special 
case of the OSD Schmidt witness. However, computing the sum of the OSCs of 
$\ket{\psi_3}\bra{\psi_3}$ one finds that $\sum_{\alpha \beta} s_\alpha s_\beta \leq 2$ 
and hence SN three is not certified by the extended CCNR criterion.

Note that there are further results on the Schmidt number detection which are at least as good 
as our first example \cite{Shuheng, Nikolai}. However, from the experimental point of view they rely on different methods.
So, giving the exact relation between those criteria and the 
OSC witnesses is left for future studies.

\section{Appendix B: Estimating bipartite entanglement monotones}
In this section, we explain how some entanglement measures can be estimated
from the OSD witness. First, we explain how the sum of the Schmidt coefficients
of a pure state can be estimated from the expectation value of an OSD witness. 
Second, we prove that if some entanglement measure is a certain function
of the sum of the Schmidt coefficients for pure states, then the value of
the measure for mixed states can be estimated from the witness value. Finally, 
for several entanglement measures a suitable functional relation with the 
sum of the Schmidt coefficients is already known from the literature \cite{observe1}, 
this then gives explicit bounds on the measures from the OSD witness.
We start with the following Lemma.

\textbf{Lemma {1}.} 
Consider a pure state $|\psi\rangle=\sum_i s_i |ii\rangle$ in the Schmidt 
decomposition with $\{s_i\}$ being its Schmidt coefficients in decreasing order. 
Then one has
\begin{equation}
\label{slemma}
\big(\sum_i s_i \big)^2 \geq \frac{\langle\psi|\XX|\psi\rangle}{\mu_1},
\end{equation}
where $\XX$  is an arbitrary Hermitian operator as in Eq.~(2) in the main 
text.

{\it Proof.} 
We start from the right-hand side of Eq.~(\ref{slemma}). Defining 
$S^A_{i'i}(k):=\langle i'|G_{k}^{A}|i\rangle$ and 
$S^B_{i'i}(k):=\langle i'|G_{k}^{B}|i\rangle$ 
one has
\begin{eqnarray}
{\langle\psi|\XX|\psi\rangle} &=& 
\sum_{ii'} s_i s_{i'}
\sum_k \sqrt{\mu_k} S^A_{i'i}(k) \sqrt{\mu_k} S^B_{i'i}(k) \nonumber\\
&\leq& 
\sum_{ii'} s_i s_{i'} 
\sqrt{
\big(\sum_k\mu_k | S^A_{i'i}(k)|^2\big) 
\big(\sum_k\mu_k | S^B_{i'i}(k)|^2\big)
} \nonumber\\
&\leq& \sum_{ii'}s_i s_{i'}  
\sqrt{ \mu_1^2 \big(\sum_k   | S^A_{i'i}(k)|^2\big)\big(\sum_k | S^B_{i'i}(k)|^2\big)} \nonumber\\
&\leq&  \mu_1 \big(\sum_i s_i \big)^2,
\end{eqnarray}
where we have used the Cauchy-Schwarz inequality,  $\mu_k \leq \mu_1$, 
and $\sum_k|S^A_{i'i}(k)|^2\leq 1$ and $\sum_k|S^B_{i'i}(k)|^2 \leq 1$. 
\hfill  $\square$

Then we can formulate the following theorem.

\textbf{Theorem {2}.} Consider an entanglement measure $E$ for bipartite states defined
via the convex roof construction and suppose that for pure states
\begin{equation}
E(|\psi\rangle)=F(\vec{s}),
\end{equation}
with $|\psi\rangle=\sum_i s_i |ii\rangle$, and $\vec{s}=(s_1, \dots, s_R)$ is the 
vector containing the Schmidt coefficients. Let us define the function $f(x)$ as
\begin{equation}
f(x)=\min_{\vec{s}}\big\{ F(\vec{s}) \; \big| \;x=\big(\sum_i {s_i}\big)^2\big\},
\end{equation}
and $co[f(x)]$ is the convex hull of $f(x)$ 
(that is the largest convex function smaller or equal to $f(x)$). 
If $co[f(x)]$ is a monotonously increasing 
function in $x$, then for an arbitrary bipartite mixed state $\varrho$ with dimension 
$m\otimes n$ ($m\leq n$), one has
\begin{equation}
E(\varrho)\geq co[f(\frac{\tr(\varrho \XX)}{\mu_1})].
\end{equation}

{\it Proof.} 
Assume that $\varrho=\sum_j p_j |\psi_j\rangle\langle \psi_j|$ is the 
optimal decomposition for $\varrho$ to achieve the infimum of 
$E(\varrho)=\inf_{p_i,|\psi_i\rangle} \sum_i p_i E(|\psi_i\rangle)$. 
Then we have
\begin{eqnarray}
E(\varrho)& = &\sum_j p_j E(|\psi_j\rangle) = \sum_j p_j F(\vec{s}_{j})
\nonumber\\
&\geq &\sum_j p_j f(x_j) \geq \sum_j p_j co[f(x_j)]\nonumber\\
&\geq& co[f(\sum_j p_j x_j)] \geq co[f(\frac{\tr(\varrho \XX)}{\mu_1})],
\end{eqnarray}
where we have used $x_j:=[\sum_i (s_i)_j]^2$, and
\begin{eqnarray}
\sum_j p_j x_j
&\geq& 
\sum_j p_j \frac{\langle\psi_j|\XX|\psi_j\rangle}{\mu_1}=\frac{\tr(\varrho \XX)}{\mu_1}
\end{eqnarray}
based on Lemma 1, and $co[f(x)]$ is a monotonously increasing convex function.   \hfill  $\square$

Finally, we can write down the lower bounds on entanglement measures:

\textbf{Corollary {3}.} For many entanglement measures $E$ the corresponding $co[f(x)]$ functions
are known \cite{observe1}. Specifically, based on Theorem 2, one can obtain lower bounds 
on the convex-roof extended negativity (CREN) $E_{\mathrm{cren}}$  \cite{CREN}, concurrence $C$ \cite{EOFde1,2qubit1,2qubit2,concurrence3,concurrence4,concurrence5}, 
G-concurrence $C_\mathrm{g}$ \cite{Gour,Fan,Uhlmann}, and geometric measure of entanglement $E_{\mathrm{gme}}$ \cite{GME1,GME2,GME21}
as follows,
\begin{subequations}
\begin{eqnarray}
E_{\mathrm{cren}}(\varrho)&\geq&\frac{1}{2}(S-1), \label{cren}\\
C(\varrho)&\geq&\sqrt{\frac{2}{m(m-1)}}(S-1), \\
C_\mathrm{g}(\varrho)&\geq&S+1-m, \\
E_{\mathrm{gme}}(\varrho)&\geq&1-\frac{1}{m^2}\Big[\sqrt{S}+\sqrt{(m-1)(m-S)}\Big]^2,
\label{gme}
\end{eqnarray}
\end{subequations}
where  $S=\max\{\tr(\varrho\XX)/\mu_1,1\}$ and $m$ is the smaller local dimension of the 
bipartite $m \times n$ system. If we choose $\XX=|\phi\rangle\langle\phi|$, the results of 
Eqs.~(\ref{cren})-(\ref{gme}) reduce to the results in Ref.~\cite{observe1}.


\section{Appendix C: Optimization of OSD witnesses}
In this appendix we explain in detail how to optimize an OSD witness 
numerically. First, we will give an algorithm that optimizes the operator 
Schmidt coefficients (OSCs) of the operator $\XX$. Then an algorithm which 
focuses on the optimization of the Schmidt operators will be introduced. 
Lastly, we will discuss the actual implementation for bipartite and multipartite
systems

\subsection{C1: Optimization with respect to the OSCs}
First, recall that the OSD witness is given by $\WW = \mu_1\openone-\XX$, 
where $\mu_1$ is the largest OSC. In order to optimize the OSD witness 
for some target state $\varrho$, the goal is to minimize the required 
visibility $p_\mathrm{crit}$, which is given by
\begin{equation}
p_\mathrm{crit} = 
\frac{\tr(\WW \sigma)}{\tr(\WW \sigma)-\tr(\WW \varrho)} \label{eq:noise}
\end{equation}
for some noise $\sigma$. In the later examples, we typically 
choose $\sigma$ to be the completely mixed state ${\openone}/{d^2}$ 
here. Using the definition of the OSD witness and writing 
$\XX = \sum_i\mu_i G_i^A\otimes G_i^B$ in its OSD, we can express 
Eq.~(\ref{eq:noise}) as
\begin{equation}
p_\mathrm{crit} = 
\frac{\mu_1-\sum_i\mu_i\tr[(G_i^A\otimes G_i^B)\sigma]}
{\sum_i\mu_i\tr[(G_i^A\otimes G_i^B)(\varrho-\sigma)]} 
= p_\mathrm{crit}(\{\mu_i\})
\end{equation}
and thus interpret it as a function of the $\{\mu_i\}$. Here, it is important to mention, that A and B no longer refer to Alice and Bob as single parties, but considering a bipartition of the multipartite system, we denote the first subsystem as Alice's part and the second one as Bob's. Note that therefore, the OSD might be asymmetric which we will address in more detail in the next subsection. As we wish
to minimize the visibility by updating the OSCs, we may apply a 
{gradient descent algorithm}. To do so, we start by computing 
the gradient of $p_\mathrm{crit}$ with respect to the OSCs yielding
\begin{equation}
\big[\vec{\nabla}_{\{\mu_i\}} p_\mathrm{crit}(\{\mu_i\})\big]_j 
= \frac{\left(\delta_{1j}-\tilde{\sigma}_j\right)
\left[\sum_i\mu_i(\tilde{\varrho}_i-\tilde{\sigma}_i)\right] 
-\left(\tilde{\varrho}_j-\tilde{\sigma}_j\right)
\left(\mu_1-\sum_i\mu_i\tilde{\sigma}_i\right)}
{\left[\sum_i\mu_i(\tilde{\varrho}_i-\tilde{\sigma}_i)\right]^2} 
\label{eq:grad}
\end{equation}
for the $j$-th entry of the gradient, where we used the short-hand 
notations $\tilde{\varrho}_i = \tr[(G_i^A\otimes G_i^B)\varrho]$ and $\tilde{\sigma}_i = \tr[(G_i^A\otimes G_i^B)\sigma]$.
Then, the witness can be improved applying the following algorithm.

\vspace{0.5cm}
\noindent
\textit{Algorithm 1 (Optimization over Schmidt coefficients):}
 \begin{itemize}
 \item Start with some input operator $\XX_\mathrm{in}$ and compute 
the OSD witness.
     \item Write $\XX_\mathrm{in} = \sum_i \mu_i G_i^A\otimes G_i^B$ 
     in the OSD.
     \item Compute the gradient $\vec{\nabla}_{\{\mu_i\}} p_\mathrm{crit}
     (\{\mu_i\})$ of the required visibility using Eq.~(\ref{eq:grad}).
     \item Update the OSCs according to the update rule 
     \begin{align}
         \mu_i\mapsto \mu_{i,\mathrm{up}} = \mu_i-\varepsilon v_i \text{\,\,with\,\,} 
         \vec{v}=\frac{\vec{\nabla}_{\{\mu_i\}} 
         p_\mathrm{crit}(\{\mu_i\})}{|\vec{\nabla}_{\{\mu_i\}} p_\mathrm{crit}(\{\mu_i\})|}.
     \end{align}
     \item Update the operator $\XX_\mathrm{in}$ using the new OSCs:
     \begin{align}
         \XX_\mathrm{up} = \sum_i \mu_{i,\mathrm{up}} G_i^A\otimes G_i^B.
     \end{align}
     \item Renormalize the updated operator $\XX_\mathrm{up}$ by fixing 
     the trace $\tr(\XX_\mathrm{up}) =\tr(\XX)$ and compute the new maximum OSC $\mu_{1,\mathrm{up}}$.
 \end{itemize}
 Iterating those steps until the algorithm converges will improve the witness. The renormalization has the advantage that the expectation values
 of the witnesses become comparable. Especially for the case $\sigma = \openone/d^2$ the expectation values for $\varrho$ are proportional to
 the required visibility $p_{\rm crit}.$
 
\subsection{C2: Optimization with respect to the Schmidt operators}

The OSD witnesses found by the previous optimization are not necessarily optimal yet, as Algorithm 1 only changes the OSCs but leaves the Schmidt operators invariant. Thus, a further algorithm, optimizing with respect to the operators, is needed. The idea is to perform an infinitesimal rotation
$O$ on the Schmidt operators $G_i^X$ ($X = A$ or $X =B$), according to
\begin{equation}
G_i^X \mapsto \sum_{k=1}^{d_X^2} O_{ik} G_k^X,
\label{eq:trafo}
\end{equation}
so it is guaranteed that the $\{G_i^X\}$ still form an ONB after the updates. 

At this point, care has to be taken about the underlying dimension. If the dimensions of Alice and Bob are not the same (say, $d_A < d_B$) and we optimize over the Schmidt operators in the larger space
($X=B$) then the operators $G_i^X$ from the OSD do not form a basis of 
the operator space, as there are only $d_A^2$ of them, and $d_B^2$ are needed. So, in this case we complete the set of operators $G_i^X$ to
form a basis by adding further orthogonal operators. These additional
operators will not directly enter the updated witness, as they do not correspond to any Schmidt coefficient. They will, however, enter in an indirect way, as in Eq.~(\ref{eq:trafo}) the operators $G_i^X$ for 
$i \leq d_A^2$ are updated with contributions from the added operators
due to the infinitesimal rotation. Finally, note that the case of unequal
dimensions occurs naturally in the multipartite scenario, as one needs
to consider non-symmetric bipartitions.

Having this in mind, we can proceed with the optimization. The 
infinitesimal rotation matrix $O$ can be written as
\begin{equation}
 O = \openone + \sum_{l=1}^{n_X} \epsilon_X^{(l)} g^{(l)}.
\label{rotmat}
\end{equation}
Here, the vector $\vec{\epsilon}_X = \varepsilon \vec{v}_X$  
(with $\Vert\vec{v}_X\Vert=1$ and $\varepsilon \ll 1$) 
denotes the rotation direction and the $g^{(l)}$ 
are the generators of the SO($N$) group, whose elements are orthogonal $N\times N$ matrices with determinant equal to one. These generators $g^{(l)}$ are a set of antisymmetric $N \times N$ matrices with only 
two nonzero entries one and minus one, there are $n_X={N(N-1)}/{2}$
of these.
For example, for $N = 3$  there are ${3(3-1)}/{2} = 3$ generators, 
given by
\begin{align}
    g^{(1)} = \begin{pmatrix}
0 & 1 & 0\\
-1 & 0 & 0\\
0 & 0 & 0
\end{pmatrix},\,\,
    g^{(2)} = \begin{pmatrix}
0 & 0 & 1\\
0 & 0 & 0\\
-1 & 0 & 0
\end{pmatrix},\,\,
    g^{(3)} = \begin{pmatrix}
0 & 0 & 0\\
0 & 0 & -1\\
0 & 1 & 0
\end{pmatrix}.
\end{align}
In the optimization, we have $N = d_X^2$ as for the transformation in 
Eq.~(\ref{eq:trafo}) $d_X^2$-dimensional matrices are needed. Hence, 
there are $n_X = {d_X^2(d_X^2-1)}/{2}$ generators in this case. 

Now we insert Eq.~(\ref{rotmat}) into~(\ref{eq:trafo}) to obtain an equation 
for the transformation of the matrices $G_i^X$ 
\begin{align}
G_i^X &\mapsto G^X_i +\varepsilon \sum_{l=1}^{n_X} \Big( v_X^{(l)}  
\sum_{k=1}^{d_X^2} g_{ik}^{(l)}G_k^X\Big).
\end{align}
Here, we wrote the infinitesimal rotation direction as 
$\vec{\epsilon}_X = \varepsilon \vec{v}_X$.

Then, the transformation can be expressed similarly to the OSC transformation in the previous part as
\begin{align}
G_i^X &\mapsto G^X_i +\varepsilon \sum_{l=1}^{n_X}  v_X^{(l)} 
(\xi^X)_i^{(l)} 
\text{,\quad with \quad}
(\xi^X)_i^{(l)} = \sum_{k=1}^{d_X^2} g_{ik}^{(l)}G_k^X, 
\label{trafo2}
\end{align}
which makes it natural to apply a gradient descent algorithm here as 
well. 

For that, we interpret $p_\mathrm{crit}$ as a function of 
the vector $\vec{v}_X$ (or $\vec{\epsilon}_X = \varepsilon\vec{v}_X$), 
which gives the direction of the matrices' rotation. In other words: 
The goal is to find an optimal rotation direction 
$\vec{\epsilon}_\mathrm{X,\,opt}= \varepsilon\vec{v}_\mathrm{X,\,opt}$ 
such that the rotated operators $\{G_i^X\}$ minimize the function 
$p_\mathrm{crit}(\{G_i^X\})=p_\mathrm{crit}(\vec{\epsilon}_X)$.

Writing the target state $\varrho$ and the separable noise $\sigma$ 
in their OSD (with OSCs $r_i$ and $s_i$ and Schmidt operators 
$K^X_i$ and $H_i^X$ respectively) and inserting the transformation 
in Eq.~(\ref{trafo2}), the visibility reads
\begin{align}
    p_\mathrm{crit} 
    = \frac{\mu_1-\tilde{S}
    -\sum_{ij}S^B_{ij}\tr[\sum_l\epsilon_A^{(l)}(\xi^A)^{(l)}_iH_j^A]}
    {\tilde{R}+\sum_{ij}R^B_{ij}\tr[\sum_l\epsilon_A^{(l)}(\xi^A)^{(l)}_iK_j^A]
    -\tilde{S}-\sum_{ij}S^B_{ij}\tr[\sum_l\epsilon_A^{(l)}(\xi^A)^{(l)}_iH_j^A]},\label{pofeps}
\end{align}
for $X = A$. Note that the equation is symmetric in $A$ and $B$, and hence
it is straightforward to write it down for $X =B$. The expressions 
$\tilde{S}$ and $\tilde{R}$ describe the following sums of constant 
terms
\begin{eqnarray}
    \tilde{S} &=&  \sum_{ij}\mu_is_j\tr(G_i^BH_j^B)\tr(G_i^AH_j^A),
    \nonumber
    \\
    \tilde{R} &=& \sum_{ij}\mu_ir_j\tr(G_i^BK_j^B)\tr(G_i^AK_j^A)
\end{eqnarray}
and further, the coefficients $S^X_{ij}$ and $R^X_{ij}$ are given by
\begin{eqnarray}
    S^X_{ij} &=& \mu_is_j\tr(G_i^XH_j^X),
    \nonumber
    \\
    R^X_{ij} &=& \mu_ir_j\tr(G_i^XK_j^X).
\end{eqnarray}
Then the gradient can be computed, where we have to take the derivative 
of $p_\mathrm{crit}$ with respect to $\epsilon_X^{(m)}$ in order to obtain 
the $m$-th entry of the $n_X$-dimensional gradient. 
Furthermore, after performing the derivation, we set the vector 
$\vec{\epsilon}_X$ to zero, since this is the starting point of 
our optimization. This yields after some computation to
\begin{align}
(\vec{\nabla}_{\vec{\epsilon}_X} p_\mathrm{crit}(\vec{\epsilon}_X))_m 
= 
\frac{\left(\tilde{R}-\tilde{S}\right)\times
\big(-\tilde{S}_{\xi^X}^{(m)}\big)
-\big(\tilde{R}_{\xi^X}^{(m)}-\tilde{S}_{\xi^X}^{(m)}\big)
\times
\left(\mu_1-\tilde{S}\right)}
{\left(\tilde{R}-\tilde{S}\right)^2},
\label{grad}
\end{align}
with $\tilde{S}_{\xi^X}^{(l)}$ and $\tilde{R}_{\xi^X}^{(l)}$ defined 
as $\tilde{S}$ and $\tilde{R}$ but replacing $G_i^X$ by $(\xi^X)_i^{(l)}$, 
where $X = A$ or $X = B$ is the party we want to update.

Next, we take a look at what actually happens to the operators $G_i^X$ 
when performing the gradient descent algorithm. Starting with the 
transformation in Eq.~(\ref{trafo2}):
\begin{align}
G_i^X &\mapsto G^X_i + \sum_{l=1}^{n_A}  \epsilon_A^{(l)}  (\xi^A)_i^{(l)},
\end{align}
the update performed in the gradient descent algorithm 
\begin{align}
    \vec{\epsilon}_X =
    - \varepsilon \vec{\nabla}_{\vec{\epsilon}_X} p_\mathrm{crit}(\vec{\epsilon}_X)
\end{align}
can be inserted 
and the final update rule for the operators $G_i^X$ is
\begin{align}
    G_i^X &\mapsto G^X_i  - \varepsilon \sum_{l=1}^{n_X}[\vec{\nabla}_{\vec{\epsilon}_X} p_\mathrm{crit}(\vec{\epsilon}_X)]^{(l)} (\xi^X)_i^{(l)}.
\end{align}
Taking a closer look at it, one can note that the update rule corresponds to the transformation in Eq.~(\ref{trafo2}) with the negative gradient as rotation direction $\vec{v}_{X} = - \vec{\nabla}_{\vec{\epsilon}_{X}} p_\mathrm{crit}(\vec{\epsilon}_{X})$.

Finally, we can summarize the optimization with respect to the Schmidt operators 
for both parties in Algorithm~2:

\vspace{0.5cm}
\noindent
\textit{Algorithm 2 (Optimization over Schmidt operators): }

\begin{itemize}
     \item Start with some input operator $\XX_\mathrm{in}$  and compute the OSD witness.
     \item Write $\XX_\mathrm{in} = \sum_i \mu_i G_i^A\otimes G_i^B$ as well as $\sigma = \sum_i s_i H_i^A\otimes H_i^B$ and $\varrho = \sum_i r_i K_i^A\otimes K_i^B$ in their OSDs.
     \item Compute the gradient $\vec{\nabla}_{\vec{\epsilon_{A}}} p_\mathrm{crit}(\vec{\epsilon}_{A})$ of the visibility.
     \item Update the Schmidt operators according to the update rule 
     \begin{align}
         G_i^{A} &\mapsto G^{A}_i  - \varepsilon \sum_{l=1}^{n_{A}}(v_{A})^{(l)} (\xi^{A})_i^{(l)} \text{\,\,with\,\,} \vec{v}=\frac{\vec{\nabla}_{\vec{\epsilon}_{A}} p_\mathrm{crit}(\vec{\epsilon}_{A})}{|\vec{\nabla}_{\vec{\epsilon}_{A}} p_\mathrm{crit}(\vec{\epsilon}_{A})|}.
     \end{align}
     \item Update the operator $\XX_\mathrm{in}$ using the new Schmidt operators as
     \begin{align}
         \XX_\mathrm{up} = \sum_i \mu_i G_{i,\mathrm{up}}^A\otimes G_i^B.
     \end{align}
     \item Renormalize the updated operator $\XX_\mathrm{up}$ and compute the new maximum OSC $\mu_{1,\mathrm{up}}.$
     \item Repeat all steps for $X=B$.
 \end{itemize}
As for Algorithm~1, repeating those steps will lead to an (improved) entanglement witness. 

\subsection{C3: Practical use of the algorithm for bipartite systems}
In this subsection, we will discuss the practical use of the algorithm. As mentioned in the main text, the best choice of the OSCs and the Schmidt operators in the bipartite case is already known. However it is sensible 
to make sure that the algorithms work as expected before extending them to 
the multipartite case.

In order to test our algorithms, we choose $\varrho = \varrho_\mathrm{UPB}$ 
to be our target state. This state is bound entangled and therefore not detected by fidelity-based witnesses \cite{faithful}. It is given by 
\cite{UPB}
\begin{equation}
    \varrho_\mathrm{UPB} = \frac{1}{4}\big(\openone-\sum_{i=0}^4\ket{\psi_i}\bra{\psi_i}\big),
\end{equation}
where the vectors
\begin{align}
    \ket{\psi_0} &= \frac{1}{\sqrt{2}}\ket{0}(\ket{0}-\ket{1}),
    \quad \quad
    \ket{\psi_1} = \frac{1}{\sqrt{2}}(\ket{0}-\ket{1})\ket{2},
    \nonumber
    \\
    \ket{\psi_2} &= \frac{1}{\sqrt{2}}\ket{2}(\ket{1}-\ket{2}),
    \quad \quad
    \ket{\psi_3} = \frac{1}{\sqrt{2}}(\ket{1}-\ket{2})\ket{0},
    \nonumber
    \\
    \ket{\psi_4} &= \frac{1}{3}(\ket{0}+\ket{1}+\ket{2})(\ket{0}+\ket{1}+
    \ket{2})
\end{align}
form an {unextendible product basis} (UPB). This means that they are 
orthogonal product vectors, but there is no further product vector orthogonal 
to all of them. 

To check whether our algorithms are able to find an entanglement witness which detects this state, we start by choosing the input operator $\XX_\mathrm{in}$ randomly and applying Algorithm~1 first.

So, we start by using the following input parameters:
\begin{eqnarray}
    \XX_\mathrm{in} &=& \XX_\mathrm{random} 
    \nonumber \\
    \varrho &=& (1-\varepsilon)\varrho_\mathrm{UPB}+\varepsilon\varrho_\mathrm{random} 
    \nonumber \\
    \sigma &=& (1-\varepsilon)\frac{\openone}{9}+\varepsilon
    \varrho_\mathrm{random},
\end{eqnarray}
with $\varepsilon = 0.0001$. The state $\varrho_\mathrm{random}$ is a random density matrix and is added to all inputs, in order break the symmetry and
to make sure that the OSD of the state is unique. Without this, it can 
happen that some OSCs are equal, then the Schmidt operators are not unique, 
and the routines for performing the Schmidt decomposition may pick 
non-hermitean Schmidt operators.

In each step of the algorithm, we compute the updated visibility and already 
after a few iterations it converges to a smaller value. Considering the 
OSCs during the updates, one observes that they become more and more equal 
with each iteration, which we expect since for the bipartite case, the 
optimal $\XX$ has only equal OSCs. Moreover, it is to note that depending on 
the input operator $\XX_\mathrm{in}$ it might be that Algorithm~1 already finds 
a witness detecting the target state. However, if the random input operator 
$\XX_\mathrm{in}$ does not have suitable Schmidt operators yet, also 
Algorithm~2 is needed in order to obtain a witness for $\varrho_\mathrm{UPB}$. 

Therefore, we now apply Algorithm~2 to find a proper witness. As the OSCs are 
already the desired ones it is natural to choose the output operator 
$\XX_\mathrm{out}$ from the first algorithm as input for the second one. Doing 
so, after a {few} iterations only, our algorithm finds a witness detecting the 
target state with visibility $p_\mathrm{crit} = 0.8908$. Besides numerical 
uncertainties this is exactly the visibility corresponding to the CCNR 
criterion. Thus, we have shown that our algorithms indeed find the 
optimal witness within the OSD witnesses.

It is to note that applying Algorithm~2 first, followed by Algorithm~1 does not necessarily lead to the optimal witness. But alternating between the two in each iteration does. Having found this, we lastly want to mention how to extend the algorithms to the multipartite case.

\subsection{C4: Extension to the multipartite case}
Considering the optimization algorithms, the first step after choosing an input 
operator $\XX_\mathrm{in}$, is performing the OSD. For two parties it is clear 
how to perform the OSD because there is only one bipartititon. However, for 
more than two parties there are more bipartitions and therefore more 
possibilities to decompose the input operator $\XX_\mathrm{in}$. Since the OSD 
witness is determined by the largest OSC with respect to all bipartitions, it 
is natural to choose the bipartition where the OSC is the maximal one. We will 
call this bipartition the {critical bipartition} in the following.

\begin{figure}[t]
\begin{center}
\includegraphics[width=0.5\columnwidth]{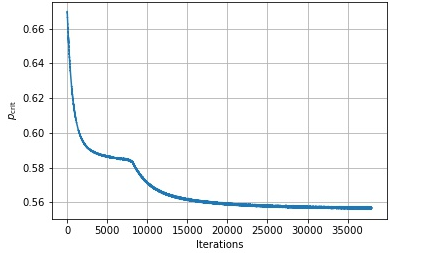}
\caption{Optimization of the required visibility for the second part of the optimization applied to the three-qubit W state $\ket{W_3}$. After applying Algorithm~1 to the fidelity witness, the output witness had a visibility greater than $p_\mathrm{fid} = 0.620$, which is the starting value for the second part of the optimization. The graph shows the optimization of the visibility when applying the two algorithms alternately to the output of the OSC optimization. After roughly 35.000 iterations, the visibility converges to the minimum $p_\mathrm{OSD} = 0.556$. For each iteration, we used the step size $\varepsilon=0.001$.} 
\label{fig1}
\end{center}
\end{figure}

Hence, we can simply extend our algorithms to the multipartite case by only 
applying them to the critical bipartition in each step. Still, it is to note 
that the critical bipartition always changes and thus now also the Schmidt 
operators change if we update the OSCs. Therefore, the optimization strategy 
in order to improve fidelity witnesses is the following:

We start by applying Algorithm~1, where we take the target state $\varrho$ 
itself as input operator $\XX_\mathrm{in}$. After some iterations the 
algorithm converges. Here it is to mention that it not necessarily converges to 
a visibility smaller than for the fidelity witness, but to a visibility 
corresponding to the optimal OSCs. In the next step it turned out to be the 
best strategy applying Algorithm~1 and Algorithm~2 alternately to the previous 
output operator $\XX_\mathrm{out}$, which finally leads to an improved witness. 
Figure~\ref{fig1} shows how the visibility behaves for the target state $\varrho = 
\ket{W_3}\bra{W_3}$, for the second part of the optimization.

We applied this procedure to several relevant states, where the most significant examples are already given in the main text (see 
Table~I in the main text). We now want to 
give the explicit forms of those states and also mention the 
results obtained for some further example states:

 First, we considered {Dicke states}. These are entangled 
 states, which were first investigated by R. H. Dicke in 1954 \cite{36}. Symmetric Dicke states are of the form
\begin{align}
    \ket{D_{k,N} }= \binom{N}{k}^{-\frac{1}{2}}\sum_i P_i(\ket{1}^{\otimes k}\otimes \ket{0}^{\otimes N-k}),
\end{align}
with $N$ denoting the number of particles and $k$ the number of excitations. The notation $P_i(\cdot)$ describes a permutation of the qubits, where the sum runs over all possible permutations. Hence, for $k=2$ excitations and $N=4$ qubits we have
\begin{align}
    \ket{D_4} = \frac{1}{\sqrt{6}}(\ket{0011}+\ket{1100}+\ket{0110}+\ket{1001}+\ket{1010}+\ket{0101}).
\end{align}
Dicke states with only $k = 1$ excitation are called {W states} \cite{k-entanglement1}. Here we found improved witnesses for
\begin{equation}
    \ket{W_3} = \frac{1}{\sqrt{3}} (\ket{001}+\ket{010}+\ket{100})
\end{equation}
and
\begin{equation}
    \ket{W_4} = \frac{1}{2} ( \ket{0001}+\ket{0010}+\ket{0100}+\ket{1000}).
\end{equation}
Note, that there also exist other works presenting witnesses specially tailored to Dicke states \cite{Bergmann}. Still, as these are quite common states it is sensible to demonstrate the performance of the OSD witnesses also on these examples.

Further, we applied our algorithms to a three-qubit uniform
{hypergraph state} given by \cite{1404.6492}:
\begin{align}
    \ket{H_3} &= \frac{1}{\sqrt{8}}( \ket{000} + \ket{001} + \ket{010} + \ket{011} + \ket{100} + \ket{101} + \ket{110} - \ket{111})\\
    &\sim \frac{1}{2}(\ket{000}+\ket{010}+\ket{100}+\ket{111}),
\end{align}
where the second expression is obtained, performing a Hadamard transformation of the third qubit.

Additionally, the {four-qubit singlet state} is given by {\cite{rev5,four-qubit}}:
\begin{align}
    \ket{\Psi_4} = \frac{1}{\sqrt{3}}(\ket{0011}+\ket{1100}-\frac{1}{2}\left(\ket{01}+\ket{10}\right)\otimes\left(\ket{01}+\ket{10}\right)).
\end{align}

We also applied our algorithms to the state $\ket{\chi}$, which is a so-called {comb monotone} state {\cite{comb}}. It is a maximally entangled state for certain entanglement measures and is defined as
\begin{align}
    \ket{\chi} = \frac{1}{\sqrt{6}}(\sqrt{2}\ket{1111}+\ket{0001}+\ket{0010}+\ket{0100}+\ket{1000}).
\end{align}
We could improve the visibility for this state from $p_\mathrm{fid} = 0.467$ to $p_\mathrm{OSD} = 0.461$. 

The last example is another maximally entangled state, the {cluster state}, which is for four qubits given by \cite{Cl41,rev5}
\begin{align}
    \ket{Cl_4} = \frac{1}{2}(\ket{0000}+\ket{1100}+\ket{0011}-\ket{1111}).
\end{align}
Here the visibility improved from $p_\mathrm{fid} = 0.467$ to $p_\mathrm{OSD} = 0.463$.
\section{Appendix D: Analytical approach}

Since the choice of the operator $\XX$ so far only relies on numerical optimization where the computational effort may grow 
with the system size, we will give now an additional 
approach to choose $\XX$ based on  analytical updates only.

First, considering the OSCs of the optimized operator $\XX$ found by the numerical algorithms, we find that for the three-qubit case, all OSCs are equal for all bipartitions: $\mu_i^\alpha = \mu$  $\forall i$ and $\forall \alpha$. This gives rise to the idea to choose $\XX$ to be a GHZ-type operator as mentioned in the main text: $\XX = \sum_{i=1}^4 \mathcal{G}_i^A\otimes \mathcal{G}_i^B \otimes \mathcal{G}_i^C$. The four-level three-particle GHZ state does not only fulfill this condition on the OSCs but also is an absolutely maximally entangled (AME) state and hence a good candidate to construct an entanglement witness.

There are two ways to interpret this structure of the operator $\XX$. For one thing, we can take it as a pure four-level three-particle (GHZ) state, where the vector of dimension four corresponds to an operator of dimension two. For the other thing this is the same as taking the projector of a pure two-level three-particle state and assigning the basis $\ket{0}\mapsto\ket{0}\bra{0}, \ket{1}\mapsto\ket{0}\bra{1}, \ket{2}\mapsto \ket{1}\bra{0}, \ket{3}\mapsto \ket{1}\bra{1}$ to it. In the following we will stay with the latter interpretation and now, having a rough idea how the structure of the optimized operator $\XX$ may look like, we introduce an update rule for the choice of the basis operators $\mathcal{G}_i^X$.

Note that in the latter interpretation, we mainly use the knowledge of the structure to define a starting point for our updates.
To derive the update rules we may use a more general decomposition of the operator $\XX$ than the one given above. Therefore, we decompose $\XX$, chosen as a projector of a GHZ state, in the following way:
\begin{align}
    \XX = \sum_{ijk} \alpha_{ijk} G_i^A \otimes G_j^B \otimes G_k^C
\end{align}
with some arbitrary hermitian orthonormal bases $\{G_i^A\}$, $\{G_i^B\}$ and $\{G_i^C\}$ and 
coefficients $\alpha_{ijk} = \tr(\XX  G_i^A \otimes G_j^B \otimes G_k^C)$. Then the OSCs are defined by the coefficients $\alpha_{ijk}$, which means that updating the operators $G_i^X, (X = A,B,C)$ leaves the OSCs unchanged. Now, we will show how to choose the $G_i^X$ such that the expectation value of $\XX$ for the target state $\varrho$ is maximized. First, consider:
\begin{align}
     \tr(\XX \rho) = \tr(\sum_{ijk} \alpha_{ijk} G_i^A \otimes G_j^B \otimes G_k^C \rho).
\end{align}
In the following, we want to perform an orthogonal transformation on the operators $G_i^X = \sum_l O_{il}G_l^X$. Starting with $X = A$, we have:
\begin{align}
    \max \tr(\XX \rho) &= \max_O  \tr(\sum_{ijk} \alpha_{ijk} \sum_{l}O_{il}G_l^A \otimes G_j^B \otimes G_k^C \rho)\\
    &= \max_O \sum_{il} O_{il} \sum_{jk} \alpha_{ijk} \tr(G_l^A \otimes G_j^B \otimes G_k^C \rho)\\
    &= \max_O \sum_{il} O_{il} F_{li}\\
    &= \max_O \tr(OF)\\
    &= \sum_i \sigma_i(F).
\end{align}
In the third step we defined $F_{li} \coloneqq \sum_{jk} \alpha_{ijk} \tr(G_l^A \otimes G_j^B \otimes G_k^C \rho)$ and in the last step we made use of the fact that the trace of some operator $F$ multiplied with an orthonormal operator $O$ is always maximized by the sum of its singular values $\sigma_i(F)$. This shows that if $F = USV^\dagger$ is the singular value decomposition of $F$, then the optimal transformation $O$ is given by $O = VU^\dagger$. We proceed analogously for $X = B$ and $X = C$, and step by step update the operators $G_i^X$. For typical examples of three qubits, after roughly 20 iterations the expectation value $\tr(\XX\rho)$ converges to a maximum. 

It is to mention that if the expectation value $\tr(\XX \rho)$ converges, the trace of the operator $\XX$ and therefore also $\tr(\XX \sigma)$ must converge. Hence, also the visibility for white noise
\begin{align}
    p_\mathrm{crit} = \frac{\mu - \tr(\XX \sigma)}{\tr(\XX \rho)-\tr(\XX \sigma)}
\end{align}
converges. Still, as we do not renormalize the operator $\XX$ here, in general it might happen that the visibility is not monotone and therefore does not converge to the optimal value, however, for all examples considered here the visibility is monotone as well and thus we find minima for the visibility, too.
Further, we stress that these updates are more general than simple local unitary transformations since already for qubits an orthogonal transformation including the identity may change the eigenvalues and the trace of the operator $\XX$. Therefore, using this method, we indeed find new entanglement witnesses (not only rotated ones).



Since the coefficients $\alpha_{ijk}$ are fixed during the optimization, the remaining question is how to choose them in a clever way, which results in the question how to choose the initial operator $\XX_0$. Indeed, taking $\XX_0 = \ket{GHZ}\bra{GHZ}$, where $\ket{GHZ} = \ket{000}+\ket{011}+ \ket{110}+\ket{101}$ and decomposing it into some arbitrary bases $\{G_i^A\}$, $\{G_i^B\}$ and $\{G_i^C\}$ to obtain the coefficients $\alpha_{ijk}$ will, after updating the operators, give an OSD witness with visibility $p_\mathrm{crit} = 0.60$ for the three-qubit W state and $p_\mathrm{crit} = 0.55$ for the hypergraph state. Thus, we can improve the fidelity witness by only performing a few analytically chosen updates on $\XX$, while for the three-qubit hypergraph state we even reach the same visibility as obtained by the numerical algorithm. 

There are two remarks in order: First, one could also choose $\ket{GHZ'} = \ket{000}+\ket{111}$ to built $\XX_0$. This gives the same result for the hypergraph state, but a worse one for the W state. The reason is that this method can easily get stuck in local optima. So it is sensible to try out different starting operators $\XX_0$. Still, taking AME states, or states with evenly distributed OSCs turn out to give good results.
Second, we can do the same procedure for more than three qubits. Taking the projector of the four- or five-qubit GHZ state as starting operator, we can improve the visibility for the four-qubit hypergraph state with one big hyperedge from $p_\mathrm{fid} = 0.87$ to $p_\mathrm{OSD,ana}=0.78$ and for five qubits from $p_\mathrm{fid} = 0.94$ to $p_\mathrm{OSD,ana}=0.92$. Hence, we introduced a method based on purely analytical steps, which allows us to find OSD witnesses with better visibility than fidelity witnesses and is also feasible for larger systems.

\section{Appendix E: Estimating multipartite entanglement monotones}
In this appendix we will show how specific entanglement monotones for genuine multiparticle
entanglement can be lower bounded via OSD witnesses. 

First, considering a bipartite entanglement measures $E$, we can define the corresponding 
genuine multipartite entanglement (GME) measure for pure states as  
\begin{equation}
E_{\mathrm{GME}}(|\psi\rangle)=\min_{\alpha} E_{\alpha}(|\psi\rangle)
\end{equation}
where $\alpha$ denotes all possible bipartitions $\alpha|\bar{\alpha}$ of $\{1,2,\cdots,N\}$ 
\cite{observe2}. This can then be extended to mixed states via the convex roof construction,
$E_{\mathrm{GME}}(\varrho)=\mathrm{inf}_{\{p_i,|\psi_i\rangle\}}\sum_{i}p_i E_{\mathrm{GME}}(|\psi_i\rangle)$, where the infimum runs over all possible decompositions of the mixed state $\varrho$. 



Suppose that the dimension of the multipartite system under the bipartition $\alpha|\bar{\alpha}$ 
is $m_\alpha\otimes n_{\bar{\alpha}}$ with $m_\alpha\leq n_{\bar{\alpha}}$. An 
arbitrarily chosen Hermitian operator $\XX$ in this  multipartite system  has its 
OSD under the bipartition $\alpha|\bar{\alpha}$ as
\begin{equation}
\label{Oalpha}
\XX=\sum_{i=1}^{m_{\alpha}^2} \mu_i^{(\alpha)} G_{i}^{(\alpha)} \otimes G_{i}^{(\bar{\alpha})},
\end{equation}
where $\{\mu_i^{(\alpha)}\}$ are the OSCs in decreasing order for the 
bipartition $\alpha|\bar{\alpha}$. As explained in the main text, if we 
define $\mu:=\max_{\alpha}\{\mu_1^{(\alpha)}\}$ then
\begin{equation}
\WW = \mu\openone-\XX
\label{eq:appendixgme}
\end{equation}
is a witness for GME. In the following, we also define $m: = \max_{\alpha} \{m_{\alpha}\}$. 
Similarly to Appendix B, one can now obtain lower bounds of GME measures based 
on the GME witness in Eq.~(\ref{eq:appendixgme}).

\textbf{Theorem {4}.} 
Consider an entanglement measure $E$ for bipartite states, such that
we have for each bipartition 
\begin{equation}
E_{\alpha}(|\psi\rangle)=F(\vec{s}^{(\alpha)}),
\end{equation}
where  $|\psi\rangle=\sum_i s_i^{(\alpha)} |ii\rangle$ is the Schmidt 
decomposition of the multiparticle state $\ket{\psi}$ for the bipartition 
$\alpha|\bar{\alpha}$. We define the function $f(x^{(\alpha)})$ as  
\begin{equation}
f(x^{(\alpha)} , m_\alpha) = 
\min_{\vec{s}^{(\alpha})} \big \{ F(\vec{s}^{(\alpha)})\;\big|\;x^{(\alpha)}=(\sum_{i=1}^{m_\alpha} s_i^{(\alpha)})^2\big\},
\end{equation}
where $m_\alpha$ denotes the dependence on the underlying dimension. Then, let 
$co[f(x^{(\alpha)},m_\alpha)]$ be the convex-hull of $f(x^{(\alpha)},m_\alpha)$ 
for $x^{(\alpha)}$. If $g(\mu,m):=f(\langle \psi|\XX|\psi\rangle/\mu,m)$ is a 
monotonously decreasing function of $\mu$ and $m$, and $co[f(x^{(\alpha)},m_\alpha)]$ 
is a monotonously increasing convex function,  then for an arbitrary $N$-partite 
mixed state $\varrho$ the bound
\begin{equation}
E_{\mathrm{GME}}(\varrho)\geq co[f(\frac{\tr(\varrho \XX)}{\mu},m)]
\end{equation}
holds.

{\it Proof.}  Assume that $\varrho=\sum_j p_j|\psi_j\rangle\langle\psi_j|$ is
the optimal decomposition for $\varrho$ to achieve the infimum of 
$E_{\mathrm{GME}}(\varrho)=\inf_{\{p_i,|\psi_i\rangle\}}\sum_i p_i E_{\mathrm{GME}}(|\psi_i\rangle)$.
Then we have $E_{\mathrm{GME}}(\varrho)=\sum_j p_j E_{\mathrm{GME}}(|\psi_j\rangle)$.
For any possible bipartition $\alpha|\bar{\alpha}$, we can calculate $E_{\mathrm{GME}}$ 
of each $|\psi_j\rangle$ as  $E_{\alpha}(|\psi_j\rangle)=F(\vec{s}^{(\alpha,j)})$, where $\vec{s}^{(\alpha,j)}$ is the vector containing the Schmidt coefficients of $|\psi_j\rangle$ under the  bipartition 
$\alpha|\bar{\alpha}$.
Then we have
\begin{eqnarray}
E_{\mathrm{GME}}(|\psi_j\rangle)
&=&\min_\alpha E_\alpha(|\psi_j\rangle)=\min_\alpha F(\vec{s}^{(\alpha, j)})
\nonumber\\
&\geq& \min_\alpha  f(x_j^{(\alpha)},m_\alpha)  \geq \min_\alpha co[f(x_j^{(\alpha)},m_\alpha)]\nonumber\\
 &\geq& co[f(\frac{\langle \psi_j|\XX|\psi_j\rangle}{\mu},m)],
\end{eqnarray}
where we have used
\begin{eqnarray}
x_j^{(\alpha)} = \big(\sum_{i=1}^{m_\alpha} \sqrt{s_i^{(\alpha,j)}}\big)^2  
\geq  
\frac{\langle \psi_j|\XX|\psi_j\rangle}{\mu_1^{(\alpha)}}  
\geq
\frac{\langle \psi_j|\XX|\psi_j\rangle}{\mu}
\end{eqnarray}
based on Lemma 1 in Appendix B and  the fact that $g(\mu,m): = f(\langle \psi|\XX|\psi\rangle/\mu,m)$ is a monotonously decreasing function of $\mu$ and $m$. Thus, we have that
\begin{eqnarray}
E_{\mathrm{GME}}(\varrho)&=&\sum_j p_j E_{\mathrm{GME}}(|\psi_j\rangle)
\geq co[f(\frac{\tr(\varrho \XX)}{\mu},m)],
\end{eqnarray}
where we have used $co[f(x^{(\alpha)},m_\alpha)]$ is a monotonously increasing convex function. \hfill
$\square$

\textbf{Corollary {5}.} Similar to Appendix B and based on Theorem 4, we can obtain 
improved experimentally accessible lower bounds for GME measures, such as CREN 
of GME $\mathcal{N}_{\mathrm{GME}}$, the concurrence of GME $C_{\mathrm{GME}}$, 
the G-concurrence of GME $G_{\mathrm{GME}}$, and the geometric measure of GME $\mathcal{G}_{\mathrm{GME}}$. The bounds are given by
\begin{subequations}
\begin{eqnarray}
&&\!\!\!\!\!\!\!\!\!\mathcal{N}_{\mathrm{GME}}\geq\frac{1}{2}({S}-1), \label{Ngme}\\
&&\!\!\!\!\!\!\!\!\!C_{\mathrm{GME}}\geq\sqrt{\frac{2}{m(m-1)}}({S}-1),\label{Cgme}\\
&&\!\!\!\!\!\!\!\!\!G_{\mathrm{GME}}\geq{S}+1-m,\label{GCgme}\\
&&\!\!\!\!\!\!\!\!\!\mathcal{G}_{\mathrm{GME}}\geq1-\frac{1}{{m}^2}\Big[\sqrt{{S}}+\sqrt{(m-1)(m-\mathcal{S})}\Big]^2, \label{Ggme}
\end{eqnarray}
\end{subequations}
where  ${S}=\max\{\tr(\varrho \XX)/\mu,1\}$. If we choose $\XX=|\phi\rangle\langle\phi|$, 
the  results of Eqs.~(\ref{Ngme})-(\ref{Ggme}) reduce to the results of Theorems 1-4 
in Ref.~\cite{observe2}, respectively. 

\twocolumngrid


\begin{thebibliography}{99}

\bibitem{arute} F. Arute, K. Arya, R. Babbush, D. Bacon, J. C. Bardin, R. Barends, R. Biswas, S. Boixo, F. G. S. L. Brandao, D. A. Buell, B. Burkett, Y. Chen, Z. Chen, B. Chiaro, R. Collins, W. Courtney, A. Dunsworth, E. Farhi, B. Foxen, A. Fowler, C. Gidney, M. Giustina, R. Graff, K. Guerin, S. Habegger, M. P. Harrigan, M. J. Hartmann, A. Ho, M. Hoffmann, T. Huang, T. S. Humble, S. V. Isakov, E. Jeffrey, Z. Jiang, D. Kafri, K. Kechedzhi, J. Kelly, P. V. Klimov, S. Knysh, A. Korotkov, F. Kostritsa, D. Landhuis, M. Lindmark, E. Lucero, D. Lyakh, S. Mandrà, J. R. McClean, M. McEwen, A. Megrant, X. Mi, K. Michielsen, M. Mohseni, J. Mutus, O. Naaman, M. Neeley, C. Neill, M. Y. Niu, E. Ostby, A. Petukhov, J. C. Platt, C. Quintana, E. G. Rieffel, P. Roushan, N. C. Rubin, D. Sank, K. J. Satzinger, V. Smelyanskiy, K. J. Sung, M. D. Trevithick, A. Vainsencher, B. Villalonga, T. White, Z. J. Yao, P. Yeh, A. Zalcman, H. Neven, and J. M. Martinis, Quantum supremacy using a programmable superconducting processor, Nature \textbf{574}, 505 (2019).

\bibitem{Pan}
S.-K. Liao, W.-Q. Cai, W.-Y. Liu, L. Zhang, Y. Li, J.-G. Ren, J. Yin, Q. Shen, Y. Cao, Z.-P. Li, F.-Z. Li, X.-W. Chen, L.-H. Sun, J.-J. Jia, J.-C. Wu, X.-J. Jiang, J.-F. Wang, Y.-M. Huang, Q. Wang, Y.-L. Zhou, L. Deng, T. Xi, L. Ma, T. Hu, Q. Zhang, Y.-A. Chen, N.-L. Liu, X.-B. Wang, Z.-C. Zhu, C.-Y. Lu, R. Shu, C.-Z. Peng, J.-Y. Wang, and J.-W. Pan, Satellite-to-ground quantum key distribution, Nature \textbf{549}, 43 (2017). 

\bibitem{nadlinger22} D. P. Nadlinger, P. Drmota, B. C. Nichol, G. Araneda, D. Main, R. Srinivas, D. M. Lucas, C. J. Ballance, K. Ivanov, E. Y.-Z. Tan, P. Sekatski, R. L. Urbanke, R. Renner, N. Sangouard, and J.-D. Bancal, Experimental quantum key distribution certified by Bell's theorem,
 Nature \textbf{607}, 682 (2022).

\bibitem{zhang22}
W. Zhang, T. van Leent, K. Redeker, R. Garthoff, R. Schwonnek, F. Fertig, S. Eppelt, W. Rosenfeld, V. Scarani, C. C.-W. Lim, and H. Weinfurter, A device-independent quantum key distribution system for distant users, 
 Nature \textbf{607}, 687 (2022).
 
 \bibitem{Lukin}
 S. Ebadi, T. T. Wang, H. Levine, A. Keesling, G. Semeghini, A. Omran, D. Bluvstein, R. Samajdar, H. Pichler, W. W. Ho, S. Choi, S. Sachdev, M. Greiner, V. Vuletić, and M. D. Lukin, 
 Quantum phases of matter on a 256-atom programmable quantum simulator, 
 Nature \textbf{595}, 227 (2021).

\bibitem{gross10}
 D. Gross, Y.-K. Liu, S. T. Flammia, S. Becker, J. Eisert,
 Quantum state tomography via compressed sensing,
 Phys. Rev. Lett. \textbf{105}, 150401 (2010).

\bibitem{cramer10}
M. Cramer, M. B. Plenio, S. T. Flammia, R. Somma, D. Gross, S. D. Bartlett, O. Landon-Cardinal, D. Poulin, and Y.-K. Liu, Efficient quantum state tomography,
Nat. Commun. {\bf 1},  149 ( 2010).

\bibitem{huang20}
H. Y. Huang, R. Kueng, and J. Preskill, Predicting
many properties of a quantum system from very few mea-
surements, Nat. Phys. {\bf 16}, 1050 (2020).

\bibitem{nguyen22}
H. C. Nguyen, J. L. Bönsel, J. Steinberg, O. Gühne,
Optimizing shadow tomography with generalized measurements,
 Phys. Rev. Lett. \textbf{129}, 220502 (2022).

\bibitem{cotler20}
 J. Cotler, F. Wilczek,
 Quantum Overlapping Tomography,
 Phys. Rev. Lett. \textbf{124}, 100401 (2020).
 
\bibitem{yu23}
 N. Yu, T.-C. Wei,
 Learning marginals suffices,
 arXiv:2303.08938.

\bibitem{brunnerbell}
N. Brunner, D. Cavalcanti, S. Pironio, V. Scarani, S. Wehner,
Bell nonlocality,
Rev. Mod. Phys. \textbf{86}, 419 (2014).
  
\bibitem{uolasteering}
R. Uola, A. C. S. Costa, H. Chau Nguyen, O. Gühne,
Quantum Steering,
Rev. Mod. Phys. \textbf{92}, 15001 (2020).

\bibitem{rev4} R. Horodecki, P. Horodecki, M. Horodecki, and K. Horodecki, Quantum entanglement,  Rev. Mod. Phys. \textbf{81}, 865 (2009).

\bibitem{rev5} O. G\"{u}hne, and G. T\'{o}th, Entanglement detection, Phys. Rep. \textbf{474}, 1 (2009). 

\bibitem{rev6} N. Friis, G. Vitagliano, M. Malik, and M. Huber, Entanglement certification from theory to experiment, Nat. Rev. Phys. \textbf{1}, 72 (2019).

\bibitem{gharibian10}
S. Gharibian, 
Strong NP-hardness of the quantum separability problem, 
Quantum Inf. Comput. \textbf{10}, 343 (2010).

\bibitem{qibook} M. A. Nielsen, and I. L. Chuang, Quantum Computation and Quantum Information, Cambridge University Press, Cambridge (2000).

\bibitem{curty04}
M. Curty, M. Lewenstein, and N. Lütkenhaus, 
Entanglement as precondition for secure quantum key distribution,
Phys. Rev.
Lett. \textbf{92}, 217903 (2004).

 \bibitem{pezze09}
 L. Pezzé and A. Smerzi, 
 Entanglement, Non-linear Dynamics, and the Heisenberg Limit 
 Phys. Rev. Lett. {\bf 102}, 100401
(2009).

\bibitem{bourennane} M. Bourennane, M. Eibl, C. Kurtsiefer, S. Gaertner, H.
Weinfurter, O. Gühne, P. Hyllus, D. Bruß, M. Lewenstein,
and A. Sanpera, Experimental Detection of Multipartite
Entanglement Using Witness Operators, Phys. Rev. Lett.
\textbf{92}, 087902 (2004).

\bibitem{gezaspin}
G. T\'oth, Entanglement witnesses in spin models,
Phys. Rev. A {\bf 71}, 010301(R) (2005).

\bibitem{terhal} 
B. M. Terhal, A Family of Indecomposable Positive Linear Maps based on Entangled Quantum States,
Linear Algebra Appl. {\bf 323}, 61 (2000).

\bibitem{optimization}
M. Lewenstein, B. Kraus, J. I. Cirac, and P. Horodecki, Optimization of entanglement witnesses,
Phys. Rev. A {\bf 62}, 052310 (2000).

\bibitem{lurccnr}
O. Gühne, M. Mechler, G. Tóth, and P. Adam, Entanglement criteria based on local uncertainty relations are strictly stronger than the computable cross norm criterion,
Phys. Rev. A {\bf 74}, 010301 (2006).

\bibitem{pianistates} 
M. Piani and C. Mora, Class of positive-partial-transpose bound entangled states associated with almost any set of pure entangled states,
Phys. Rev. A \textbf{75}, 012305 (2007).

\bibitem{dariusreview}
D. Chruściński and G. Sarbicki, Entanglement witnesses: construction, analysis and classification,
J. Phys. A: Math. Theor. {\bf 47}, 483001 (2014).

\bibitem{HW}A. Asadian, P. Erker, M. Huber, and C. Klöckl, Heisenberg-Weyl Observables: Bloch vectors in phase space, Phys. Rev. A {\bf 94}, 010301(R) (2016). 

\bibitem{SIC} J. Shang, A. Asadian, H. Zhu, and O. Gühne, Enhanced entanglement criterion via symmetric informationally complete measurements, 
Phys. Rev. A {\bf 98}, 022309 (2018).

\bibitem{weilenmann20} 
M. Weilenmann, B. Dive, D. Trillo, E. A. Aguilar, and M. Navascués, 
Entanglement Detection beyond Measuring Fidelities,
Phys. Rev. Lett. {\bf 124}, 200502 (2020); 
Erratum: Phys. Rev. Lett. {\bf 125}, 159903 (2020).

\bibitem{faithful} O. G\"{u}hne, Y. Mao, and X.-D. Yu, Geometry of Faithful Entanglement, Phys. Rev. Lett. \textbf{126}, 140503 (2021).

\bibitem{Nph1} X.-L. Wang, L.-K. Chen, W. Li, H.-L. Huang, C. Liu, C. Chen, Y.-H. Luo, Z.-E. Su, D. Wu, Z.-D. Li, H. Lu, Y. Hu, X. Jiang, C.-Z. Peng, L. Li, N.-L. Liu, Y.-A. Chen, C.-Y. Lu, and J.-W. Pan, Experimental ten-photon entanglement, Phys. Rev. Lett. \textbf{117}, 210502 (2016).

\bibitem{Nph2} X.-C. Yao, T.-X. Wang, P. Xu, H. Lu, G.-S Pan, X.-H. Bao, C.-Z. Peng, C.-Y. Lu, Y.-A. Chen, J.-W. Pan, Observation of eight-photon entanglement, Nature Photon. \textbf{6}, 225-228 (2012). 

\bibitem{Nph3} Y.-F. Huang, B.-H. Liu, L. Peng, Y.-H. Li, L. Li, C.-F. Li and G.-C. Guo, Experimental generation of an eight-photon Greenberger-Horne-Zeilinger state, Nature Commun. \textbf{2}, 546 (2011).

\bibitem{oS} M. A. Nielsen, Quantum information theory,Ph.D. thesis, University of New Mexico, Albuquerque, (1998).

\bibitem{oS2} M. A. Nielsen, C. M. Dawson, J. L. Dodd, A. Gilchrist, D. Mortimer, T. J. Osborne, M. J. Bremner, A. W. Harrow, and A. Hines, Quantum dynamics as a physical resource, Phys. Rev. A \textbf{67}, 052301 (2003).

\bibitem{cariello} D. Cariello, Separability for Weak Irreducible matrices, Quantum Inf. Comput. \textbf{14}, 1308 (2014).

\bibitem{cariello2} D. Cariello, Analytical techniques on multilinear problems, PhD thesis, Universidad Complutense de Madrid, Madrid, (2017).

\bibitem{block} N. Johnston, http://www.njohnston.ca/2014/06/whatthe-operator-schmidt-decompositiontells-us-about-entanglement/.

\bibitem{innsbruck} G. De las Cuevas, T. Drescher, and T. Netzer, Separability for mixed states with operator Schmidt rank two, Quantum \textbf{3}, 203 (2019).

\bibitem{rescentOSD} R. Mansuroglu, A. Adil, M. J. Hartmann, Z. Holmes, and A. T. Sornborger, Quantum Tensor Product Decomposition from Choi State Tomography, arXiv:2402.05018v1

\bibitem{Eisert} J. Eisert, and H. J. Briegel, Schmidt measure as a tool for quantifying multiparticle entanglement, Phys. Rev. A \textbf{64}, 022306 (2001).

\bibitem{SN} 
B. M. Terhal, and P. Horodecki, 
Schmidt number for density matrices, 
Phys. Rev. A \textbf{61}, 040301(R) (2000).

\bibitem{schmidtwit} 
A. Sanpera, D. Bru{\ss}, and M. Lewenstein, 
Schmidt number witnesses and bound entanglement, 
Phys. Rev. A \textbf{63}, 050301(R) (2001).




\bibitem{CCNRschmidt} N. Johnston, and D. W. Kribs, Duality of Entanglement Norms, Houston J. Math. \textbf{41(3)}, 831-847 (2015). 

\bibitem{ppt} M. Horodecki, P. Horodecki, and R. Horodecki, Separability of mixed states: necessary and sufficient conditions, Phys. Lett. A \textbf{223}, 1-8 (1996).

\bibitem{ppt1} A. Peres, Separability criterion for density matrices, Phys. Rev. Lett. \textbf{77}, 1413 (1996). 

\bibitem{negativity} K. \.{Z}yczkowski, P. Horodecki, A. Sanpera, and M. Lewenstein, Volume of the set of separable states, Phys. Rev. A \textbf{58}, 883 (1998).

\bibitem{negativity21} G. Vidal, and R. F. Werner, Computable measure of entanglement, Phys. Rev. A \textbf{65}, 032314 (2002). 

\bibitem{CREN} S. Lee, D.-P. Chi, S.-D. Oh, and J. Kim, Convex-roof extended negativity as an entanglement measure for bipartite quantum systems, Phys. Rev. A \textbf{68}, 062304 (2003). 

\bibitem{EOFde1} C. H. Bennett, D. P. DiVincenzo, J. A. Smolin, and W. K. Wootters, Mixed-state entanglement and quantum error correction, Phys. Rev. A \textbf{54}, 3824-3851 (1996).

\bibitem{2qubit1} S. Hill, and W. K. Wootters, Entanglement of a Pair of Quantum Bits, Phys. Rev. Lett. \textbf{78}, 5022 (1997).

\bibitem{2qubit2} W. K. Wootters, Entanglement of formation of an arbitrary state of two qubits, Phys. Rev. Lett. \textbf{80}, 2245 (1998).

\bibitem{concurrence3} K. Audenaert, F. Verstraete, and B. De Moor, Variational characterizations of separability and entanglement of formation, Phys. Rev. A \textbf{64}, 052304 (2001).

\bibitem{concurrence4} P. Rungta, V. Buzek, C. M. Caves, M. Hillery, and G. J. Milburn, Universal state inversion and concurrence in arbitrary dimensions, Phys. Rev. A \textbf{64}, 042315 (2001). 

\bibitem{concurrence5} P. Badziag, P. Deuar, M. Horodecki, P. Horodecki, and R. Horodecki, Concurrence in arbitrary dimensions, J. Mod. Opt. \textbf{49}, 1289 (2002).

\bibitem{Uhlmann} A. Uhlmann, Roofs and convexity, Entropy \textbf{12}, 1799 (2010).

\bibitem{Gour} G. Gour, Family of concurrence monotones and its applications, Phys. Rev. A \textbf{71}, 012318 (2005).

\bibitem{Fan} H. Fan, K. Matsumoto, and H. Imai, Quantify entanglement by concurrence hierarchy, J. Phys. A \textbf{36}, 4151 (2003).

\bibitem{GME1} H. Barnum, and N. Linden, Monotones and invariants for multi-particle quantum states, J. Phys. A \textbf{34}, 6787 (2001). 

\bibitem{GME2} T.-C. Wei, and P. M. Goldbart, Geometric measure of entanglement and applications to bipartite and multipartite quantum states, Phys. Rev. A \textbf{68}, 042307 (2003).

\bibitem{GME21} T.-C. Wei, J. B. Altepeter, P. M. Goldbart, and W. J. Munro, Measures of entanglement in multipartite bound entangled states, Phys. Rev. A \textbf{70}, 022322 (2004). 

\bibitem{chen} K. Chen, S. Albeverio, and S.-M. Fei, Concurrence of Arbitrary Dimensional Bipartite Quantum States, Phys. Rev. Lett. \textbf{95}, 040504 (2005).

\bibitem{CCNR1} O. Rudolph, Further Results on the Cross Norm Criterion for Separability, Quantum Inf. Process. \textbf{4}, 219 (2005).

\bibitem{CCNR2} K. Chen and L.-A. Wu, A matrix realignment method for recognizing entanglement, Quantum Inf. Comput. \textbf{3}, 193 (2003).

\bibitem{Lancien} C. Lancien, O. G\"uhne, R. Sengupta, and M. Huber, Relaxations of separability in multipartite systems: Semidefinite programs, witnesses and volumes, J.  Phys. A: Math. Theor. \textbf{48}, 505302 (2015).

\bibitem{acinprl2001} A. Acín, D. Bruß, M. Lewenstein, and A. Sanpera, Classification of Mixed Three-Qubit States, Phys. Rev. Lett. \textbf{87}, 040401 (2001).

\bibitem{1404.6492} O. Gühne, M. Cuquet, F. E. S. Steinhoff, T. Moroder, M. Rossi, D. Bruß, B. Kraus, and C. Macchiavello, Entanglement and nonclassical properties of hypergraph states, J.  Phys. A: Math. Theor. \textbf{47}, 335303 (2014).

\bibitem{bounds15} O. G\"{u}hne, and M. Seevinck, Separability criteria for genuine multiparticle entanglement, New J. Phys. \textbf{12}, 053002 (2010). 

\bibitem{chengjieCV} C. Zhang, S. Yu, Q. Chen, and C. H. Oh, Detecting and Estimating Continuous-Variable Entanglement by Local Orthogonal Observables, Phys. Rev. Lett. \textbf{111}, 190501 (2013).

\bibitem{Jungnitsch} B. Jungnitsch, T. Moroder, and O. Gühne, Taming Multiparticle Entanglement, Phys. Rev. Lett. {\bf 106}, 190502 (2011).

\bibitem{teleportation} N. Ganguly, S. Adhikari, A. S. Majumdar, and J. Chatterjee, Entanglement Witness Operator for Quantum Teleportation, Phys. Rev. Lett. {\bf 107}, 270501  (2011).

\bibitem{distillability} R. O. Vianna, and A. C. Doherty, Distillability of Werner states using entanglement witnesses and robust semidefinite programs, Phys. Rev. A {\bf 74}, 052306 (2006).

\bibitem{Huber} M. Huber, and J. I. de Vicente, Structure of Multidimensional Entanglement in Multipartite Systems, Phys. Rev. Lett. \textbf{110}, 030501 (2013).

\bibitem{SVec} L. Bulla, M. Pivoluska, K. Hjorth, O. Kohout, J. Lang, S. Ecker, S. P. Neumann, J. Bittermann, R. Kindler, M. Huber, M. Bohmann, and R. Ursin, Nonlocal Temporal Interferometry for Highly Resilient Free-Space Quantum Communication, Phys. Rev. X \textbf{13}, 021001 (2023).

\bibitem{flammia} S. T. Flammia, and Y.-K. Liu, Direct Fidelity Estimation from Few Pauli Measurements, Phys. Rev. Lett. {\bf 106}, 230501 (2011). 

\bibitem{morelli} S. Morelli, H. Yamasaki, M. Huber, and A. Tavakoli, Entanglement Detection with Imprecise Measurements, Phys. Rev. Lett. \textbf{128}, 250501 (2022).



\bibitem{ma} S. Denker, Characterizing multiparticle entanglement using the Schmidt decomposition of operators, Springer 2023

\bibitem{Shuheng} S. Liu, Q. He, M. Huber, O. G\"{u}hne, and G. Vitagliano, Characterizing Entanglement Dimensionality from Randomized Measurements, PRX Quantum \textbf{4}, 020324 (2023).

\bibitem{Nikolai} N. Wyderka and A. Ketterer, Probing the Geometry of Correlation Matrices with Randomized Measurements, PRX Quantum \textbf{4}, 020325 (2023).

\bibitem{observe1} C. Zhang, S. Yu, Q. Chen, H. Yuan,  and C. H. Oh, Evaluation of entanglement measures by a single observable, Phys. Rev. A \textbf{94}, 042325 (2016).

\bibitem{UPB} C. H. Bennett, D. P. DiVincenzo, T. Mor, P. W. Shor, J. A. Smolin and B. M. Terhal, Unextendible Product Bases and Bound Entanglement, Phys. Rev. Lett. \textbf{82}, 5385 (1999).

\bibitem{36} R. H. Dicke, Coherence in Spontaneous Radiation Processes, Phys. Rev. {\bf 93}, 99 (1954).

\bibitem{k-entanglement1} W. D\"{u}r, G. Vidal, and J. I. Cirac, Three qubits can be entangled in two inequivalent ways, Phys. Rev. A \textbf{62}, 062314 (2000).

\bibitem{Bergmann}  M. Bergmann, and O. G\"{u}hne, Entanglement criteria for Dicke states, J. Phys. A: Math. Theor. \textbf{46}, 385304 (2013).

\bibitem{four-qubit} H. Weinfurter, and M. Żukowski, Four-photon entanglement from down-conversion, Phys. Rev. A {\bf 64}, 010102(R)  (2001).

\bibitem{comb} B. Kraus, Entanglement properties of quantum states and quantum operations, PhD thesis, Universität zu Innsbruck (2003).

\bibitem{Cl41} H. J. Briegel, and R. Raussendorf, Persistent Entanglement in Arrays of Interacting Particles, Phys. Rev. Lett. {\bf 86}, 910 (2001).

\bibitem{observe2} Y. Dai, Y. Dong, Z. Xu, We. You, C. Zhang, and O. G\"{u}hne, Experimentally Accessible Lower Bounds for Genuine Multipartite Entanglement and Coherence Measures, Phys. Rev. Applied \textbf{13}, 054022 (2020).

\end{thebibliography}
\end{document}